\documentclass{aa}
\usepackage{txfonts}
\pdfoutput=1
\usepackage{epsfig,appendix,array,rotating,pdflscape,array,longtable,colortbl,appendix,footnote,cancel}
\usepackage{multirow,url}
\usepackage{booktabs}
\usepackage{lscape, amsmath} 
\usepackage{paralist}
\usepackage{relsize}
\usepackage{amssymb}
\usepackage{xtab}
\usepackage{hyperref}
\usepackage[normalem]{ulem}
\usepackage{natbib}
\usepackage[rightcaption]{sidecap}
\bibpunct{(}{)}{;}{a}{}{,} 

\newcommand{\hi}{H\,{\sc i}}

\newcommand{\prim}{$^{\prime}$}
\newcommand{\prin}{$^{\prime\prime}$}

\newcommand{\km}{km\,s$^{-1}$}
\newcommand{\degree}{$^{\circ}$}

\newcommand{\msolar}{M$_{\odot}$}

\newcommand{\mhi}{$M({\mathrm {HI}})$}

\newcommand {\apgt} {\ {\raise-.5ex\hbox{$\buildrel>\over\sim$}}\ }
\newcommand {\aplt} {\ {\raise-.5ex\hbox{$\buildrel<\over\sim$}}\ }

\setlongtables

\begin{document}

\label{firstpage}


\title{High-resolution HI mapping of nearby {extremely metal-poor} blue compact dwarf  galaxies}

\author{Tom\ C.\ Scott$^{1}$\thanks{E-mail: tom.scott@astro.up.pt (TCS)}
\and Elias \ Brinks$^{2}$
\and Chandreyee Sengupta$^{3}$
\and Patricio  Lagos$^{1}$} 

\institute{Institute of Astrophysics and Space Sciences (IA), Rua das Estrelas, 4150--762 Porto, Portugal
\and Centre for Astrophysics Research, University of Hertfordshire, College Lane, Hatfield, AL10 9AB, UK 
\and Purple Mountain Observatory (CAS), No. 10 Yuanhua Road, Qixia District, Nanjing 210034, China}

\date{Received <date> /
Accepted <date>}


\abstract 
{}
{Optical observations of blue compact dwarf galaxies (BCDs) show they typically have high specific star formation rates \textcolor{black}{(sSFRs)} and low metallicites. A subset of these galaxies (those with the lowest gas phase metallicities) display cometary optical morphologies similar to those found at high redshift. Whether this combination of properties predominantly arises from interactions with neighbours or via accretion from the cosmic web, or is indeed due to something else, remains unclear. Our aim is to use high-resolution \hi\ mapping to gain insights into the processes driving the observed properties of a sample of extremely metal-poor (XMP) BCDs.}  
{We present Very Large Array {B-- and C--configuration} \hi\  mapping of {the} four BCDs of our sample. For three of the targeted BCDs, we also detected and mapped the \hi\ in their nearby companions.}
{In these three cases, there is \hi\ morphological and kinematic evidence of a recent flyby interaction between the BCD and a nearby companion galaxy.  The \hi\ evidence for recent interactions for these three BCDs is corroborated by our analysis of the tidal forces exerted on the BCDs by companions with available spectroscopic redshifts. In one of these cases, J0204--1009, we obtain sufficient spatial resolution to determine that the BCD is dominated by  dark matter \textcolor{black}{(DM)} and estimate its DM halo mass to be in the range of {1.2 $\times$ 10$^{11}$ to 5.2 $\times$ 10$^{11}$ } \msolar.  However, it is the most isolated BCD  in our small sample, J0301--0052, that shows {one of the most  asymmetric \hi\ morphologies}. J0301--0052 has a similar cometary \hi\ morphology to its optical morphology, although the \hi\ {column density maximum is projected at the end of the optical} tail.}
{
Our \hi\ observations suggest that J0301--0052 may be undergoing a merger, while the other members of our BCD  sample show evidence of a recent tidal interaction with a near neighbour. While our selection criteria favour BCDs with companions, our results are consistent with previous literature showing that most BCDs are associated with either mild tidal interactions or mergers.}
{}


\maketitle
\titlerunning{VLA mapping low–metallicity BCDs}

\section{introduction}
\label{into}
Star-forming blue compact dwarfs (BCDs) with M$_B$ $\geq$ -18 normally display high gas-to-stellar-mass fractions, high specific star formation rates (sSFRs), {and low metallicities (Z$_{\odot}$/40 $\lesssim$ Z $\lesssim$ Z$_{\odot}$/3)} in comparison with other dwarf galaxies \citep{thuan81,thuan83,gil03,geha06}. Based on the analysis of automatic plate measuring machine catalogues, \cite{telles00} concluded that star formation \textcolor{black}{(SF)} in {HII/BCD galaxies} is not triggered by strong tidal interactions. However, a number of other authors have determined that the enhanced \textcolor{black}{star formation rates (SFRs)} in BCDs are associated with weaker interaction signatures or gas-rich dwarf--dwarf mergers \citep[e.g.][]{taylor97,ostlin01,bekki08,martinez12}. \cite{Pustilnik01} concluded that $\sim$ 80\%  of the starbursts in their sample of 86 BCDs are attributable to tidal interactions, including weak interactions,  or low-mass mergers.  Resolved \hi\ studies of the connection between SFR and \hi\ in nearby  BCDs, using the National Radio Astronomy Observatory's {\textit{Karl G. Jansky} Very Large Array\footnote{The National Radio Astronomy Observatory is a facility of the National Science Foundation operated under cooperative agreement by Associated Universities, Inc.} (VLA)},  found evidence for the infall of gas clouds and/or mergers consistent with the interaction hypothesis, according to which their recent \textcolor{black}{SF} enhancement is triggered by recent interactions \citep{bravo04,ekta08,Ashley17}.

As galaxies evolve, their stellar mass and gas-phase metal abundance are expected to increase. As a result, metal-poor, low-stellar-mass galaxies are predicted to be common at high redshift. Conversely, few galaxies with this combination of properties are observed at low redshift. Only a few hundred local, extremely metal-poor galaxies with oxygen abundances of less than  10\% of solar, 12 + log(O/H) $\lesssim$ 7.60 (XMP galaxies), have been identified to date \citep[e.g.][]{morales11,guseva17}. Typically,  XMPs are BCDs displaying strong optical emission lines, blue colours, high surface brightness, low luminosity, compact morphologies, and faint blue optical continuum \citep[e.g.][]{kunth00,bergvall02,noeke03,gil05,lagos11,mamon12, micheva13}.

Three-quarters of XMPs also display cometary optical morphologies \citep{papaderos08,morales11}, which is an unusual morphology amongst low-redshift UV-bright star-forming galaxies \citep{elmegreen12}. Several mechanisms have been proposed to explain this cometary or tadpole morphology, including gravitational triggering from accretion of a low-mass companion \citep{straughn06}, ram pressure \citep{elmegreen10,gavazzi01c}, or the accretion of nearly pristine cold gas from the cosmic web \citep{sanchez15,Kurapati2024a}.

There has been much discussion in the literature about the mechanism(s) responsible for triggering the observed high \textcolor{black}{SFRs} in XMP BCDs and their intriguing stellar and nebular morphologies, which resemble galactic discs in formation \citep{BournaudElmegreen2009}. For example, an \hi\ interferometric study of six of the most metal-deficient star bursting dwarfs, most of which are BCDs, showed evidence of a recent interaction and/or merger in their \hi\ morphologies and kinematics  \citep{ekta08}. One possible mechanism is that the sSFR becomes enhanced following the infall of a significant mass of metal-poor gas, as in the case of the BCD UM\,461 \citep{lagos18}. Another possibility is that a recent interaction has driven gas to the galaxy centre and has triggered a burst of \textcolor{black}{SF}; see for example {\cite{Montuori10}}. This is the same process inferred for optical emission in Seyfert galaxies \citep{dahari84}.  However, the cause of the low metallicity of BCDs in this second scenario remains difficult to explain. One explanation for the low metallicity could be the expulsion of heavy elements by starburst-driven outflows  \citep[e.g.][]{ott03}. This mechanism is supported for BCDs in low-density environments by the finding that dwarf galaxy metallicity increases with proximity to the cores of clusters with higher intracluster-medium (ICM) core densities. In these {clusters, expulsion} of heavy elements from dwarfs during starbursts 
could have been  limited by the combined effect of ambient ICM pressure and ram pressure, thereby increasing their metallicity compared to dwarfs in less dense environments \citep{petrop12}.
Yet another possibility is that these are primordial objects that failed, until recently, to convert significant amounts of their cold gas into stars \citep{bergvall95,noeske07}. However, BCDs generally contain older stellar populations, which is inconsistent with this last scenario \citep{papaderos96,ostlin00}. Finally,
{most dwarf galaxies tend to show flat abundance (O/H, N/O) gradients \cite[e.g.][]{kobulnickyskillman97,leeskillman04,lagos09,lagos12,lagospapaderos13}, suggesting  rapid and efficient dispersion and/or mixing of metals in the interstellar medium (ISM) by expanding starburst-driven superbubbles and/or external gas infall. However, in \cite{lagos14,lagos16,lagos18} we found indications of variation of oxygen abundance in the ISM of some XMP BCDs. \cite{sanchez15} interpret this metallicity variation (with the lowest metallicity value localised {in}  the highest SFR region) as a sign of external metal-poor gas accretion from the cosmic web.} 

To date, there have been relatively few \hi\ interferometric observations of XMP BCDs; but where these exist, the \hi\ spatial distributions and velocity profiles are found to be distorted, suggesting either the infall of {metal-poor} gas \citep{ekta10} or recent interactions with either external galaxies \citep{ekta08} or an intergroup or intracluster medium.
\textcolor{black}{\cite{Kurapati2024a}, using} GMRT \hi\ observations, found that their sample of five XMP dwarf galaxies in voids display an overall disturbed \hi\ morphology and velocity fields, suggesting cold gas accretion or minor interactions with nearby dark galaxies. These authors propose that cold gas accretion in those systems may be connected to the presence of small filaments that form the substructure of voids \citep{aragon13}; s{ee} also \cite{kurapati2024}.

With the questions in the literature regarding the origin of the current \textcolor{black}{SF}, {optical} morphology, and metal-poor gas in mind, we selected four metal-poor XMP BCDs  from the \cite{filho13} sample to map their \hi\ with the VLA {using the} B- and C-configurations. We aim to use the resolved \hi\ maps to investigate whether one or a combination of the proposed mechanisms are consistent with the properties of these galaxies. Specifically, we examine the \hi\ morphologies and kinematics from our observations for evidence that interactions could be responsible for the triggering of their high sSFRs and/or optical cometary morphologies. 

Section \ref{sample} describes the sample, with Sect. \ref{obs} summarising the observations and data reduction. The observational results are presented in Sect. \ref{results}. A discussion follows in Sect. \ref{discussion} with concluding comments in Sect. \ref{concl}.  For consistency with previous observations, we adopt distances to the targets and spatial scales taken from the NASA/IPAC Extragalactic Database (NED), which are listed in Table \ref{table1}. All positions referred to throughout this paper are in J2000.0 and we adopt the following cosmological parameters: H$_0$ = 69.6 \km\ Mpc$^{-1}$, $\Omega_M$ = 0.286, and  $\Omega_{vac}$ = 0.714 

\section{The sample}
\label{sample}
{\cite{filho13} used the Effelsberg 100m single-dish telescope to observe a {subset of 29 out of 140 local XMP galaxies selected from 
\textcolor{black}{the Sloan Digital Sky Survey (SDSS) DR7} by \cite{morales11}. The Effelsberg full width at half power (FWHP) beam at 21cm is $\sim$ 9 arcmin. } 
{{We limited our sample selection to the {10}  of these 29 galaxies with an {Effelsberg} \hi\ detection.} The sample was further refined based on {detectability with the VLA as well as} cometary and/or disturbed optical morphology and {strength of \hi\ in their  {Effelsberg} spectrum. Our final selection contains } four XMP BCD galaxies: J0204--1009, J0301--0052, J0315--0024, and J2053+0039.}
{ Detailed descriptions of these systems are provided below.}
Basic optical {properties and results from single-dish \hi\ observations for}  the four targets are set out in Table \ref{table1} {and their DECaLS\footnote{The Dark Energy Camera Legacy Survey \citep[DECaLS;][]{dey19}.} false colour ($g$, $r$, $z$)  images are shown in Appendix \ref{appendix_A}. }}


\begin{table*}
\footnotesize
\begin{minipage}{180mm}
\caption{XMP-BCD targets: Selected single-dish HI and optical properties}
\label{table1}
\begin{tabular}{@{}lcccccccccc@{}}
\hline
ID&RA\footnote{RA, DEC position from SDSS.}&DEC&M$_*$\footnote{{M$_*$ from \citep{filho13}. }} &{12+log(O/H)}\footnote{12+log(O/H) from \citep{filho13}. }  &S$_{HI}$\footnote{Integrated \hi\ flux densities from the Effelsberg 100m  single-- dish radio telescope \citep{filho13}. }&V$_{HI}$\footnote{{V$_{HI}$}\ from the Effelsberg 100m  single-dish radio telescope \citep{filho13}. }&\hi\ W$_{50}$\footnote{W$_{50}$ from the Effelsberg 100m  single-dish radio telescope \citep{filho13}.}&\mhi\footnote{\mhi\ from the Effelsberg 100m  single-dish radio telescope \citep{filho13}.}&Distance\footnote{Distance from NED.}&Spatial scale\footnote{Spatial scale from NED.}\\
& [hms]&[dms]&[10$^8$\msolar]&&[Jy \km] &[\km]&[\km]&[10$^9$\msolar]&[Mpc]&[{pc/arcsec}]\\
 \hline
J0204$-$1009& 02 04 25.6&--10 09 34.99&0.2&7.36&9.9$\pm$0.7&1906&112&1.5 &27.12&{131.5} \\
J0301$-$0052  &03 01 49.0&--00 52 57.38&0.4&7.52&2.1$\pm$0.7&2108&110&0.4&31.22&{151.3}\\
J0315$-$0024&03 15 59.9&--00 24 26.06 &0.6&7.41&1.3$\pm$0.6&6650&92 &2.5&96.96&{470.0} \\ 
J2053$+$0039&20 53 12.6&+00 39 14.04&0.2&7.22&1.6$\pm$0.7&3907&53&1.2&51.61&{298.7}\\

 \hline
\end{tabular}
\end{minipage}
\end{table*}

\subsection{J0204--1009}
\label{sample_J0204s}
\begin{figure*}
\begin{center}
\includegraphics[ angle=0,scale=0.6] {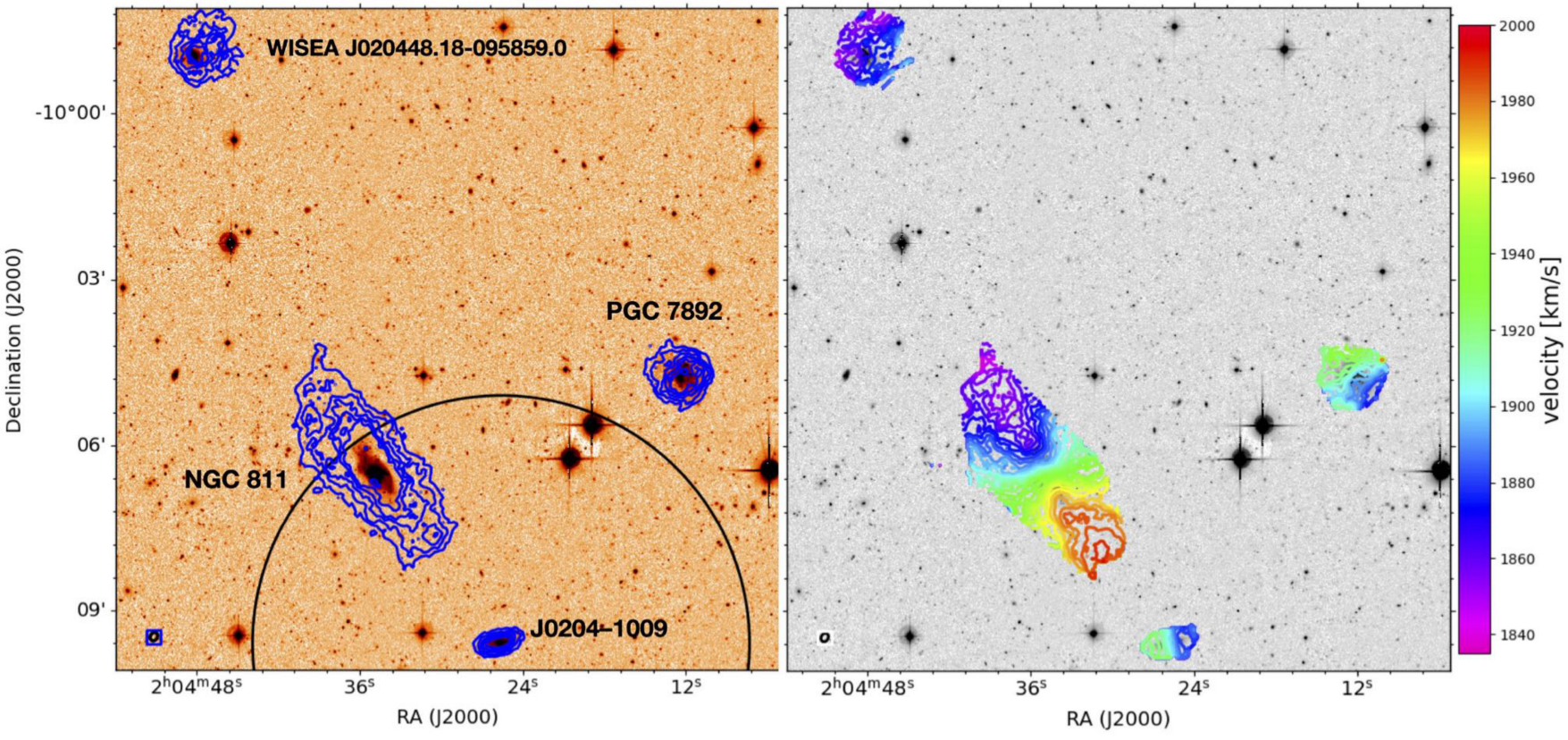}
\vspace{0.01cm}
\caption{\textbf{\textit{Left:}}  \textcolor{black}{J0204-1009 field}.
\hi\ detected in the BCD target J0204-1009 and  its three nearest neighbours, NGC\,811,  PGC\,7892, and WISEA\,J020448.18-095859.0, shown with blue contours.The VLA BC-configuration \hi\ contours are at column densities of 4.1, 8.2, 12.3, 16.5, 20.6, and 41.1 $\times$ 10$^{20}$ atoms cm$^{-2}$, with the first contour at 3 $\sigma$. The IDs of the galaxies are marked on a CFHT $g$-band image. The small ellipse in the bottom left corner indicates the size and orientation of the VLA BC-configuration 9.5\prin\ $\times$ 7.6\prin\ FWHP synthesised beam.  The large circle shows the  Effelsberg 9\prim\ FWHP beam. \textbf{\textit{Right:}} \hi\ velocity field with contours separated by 5 \km, with velocities shown by the colour scale.}
\label{fig_1}
\end{center}
\end{figure*}

J0204--1009 (also known as PGC\,007896 or KUG\,0201--103),V$_{opt}$ = 1902 \km, has the lowest measured stellar mass, 0.2 $\times$  10$^8$ \msolar, of the  seven catalogued members of the PGC\,7998 group (V$_{opt}$ = 1915 \km,  $\sigma$ = 38 \km; \citealt{kourkchi17}). The two nearest group members in projection to J0204--1009 are NGC\, 811 (V$_{opt}$ = 1923 \km) and   PGC\,7892 (V$_{opt}$ = 1923 \km); see Fig. \ref{fig_1}. The projected separation of these two group members from J0204--1009 is 3.2 arcmin (30 kpc) and 5.9 arcmin (46 kpc), respectively. The small V$_{opt}$ range of 21 \km\ and the projected separation of $\leq$ 46 kpc, together with asymmetries in their optical Canada France Hawaii Telescope (CFHT) $g$-band images, suggest a recent interaction between J0204--1009, NGC\,811, PGC\,007892, and/or other group members is highly likely.


\subsection{J0301--0052}
\label{sample_J0301s}
J0301--0052 (SDSS J030149.02-005257.3), as shown in Fig. \ref{fig_2}, has a V$_{opt}$ = 2194 \km. The SDSS $g$-band image for J0301--0052 shows this BCD has a distinct cometary morphology with the tail oriented to the SW. The cometary morphology is even more visible in the higher quality  DECaLS $g$-band {and colour  (Fig. \ref{fig_a1})} images. {DECaLS $r$- and $z$-band images show the same cometary morphology, indicating the tail includes an older stellar population.} 
A NED search within a 60 arcmin (545 kpc) radius and $\pm$ 200 \km\ of J0301--0052  revealed no galaxies. Searching the SDSS online database with similar parameters also revealed no optical candidates.

\subsection{J0315-0024}
\label{sample_J0315s}
\begin{figure*}
\begin{center}
\includegraphics[ angle=0,scale=0.25] {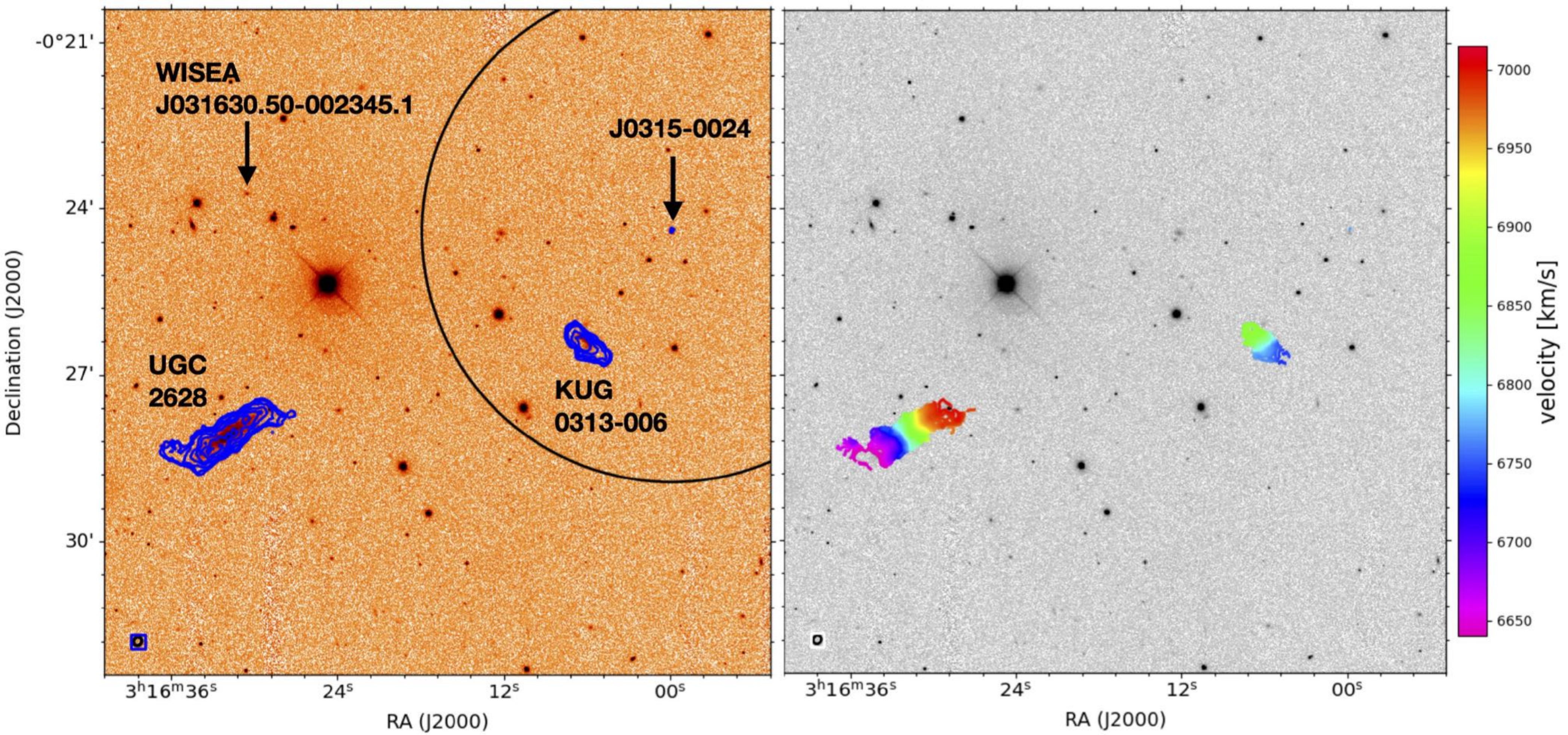}
\vspace{0.01cm}
\caption{\textbf{\textit{Left:}} \textcolor{black}{J0315--0024 field.}
\hi\ detected in the BCD target J0315--0024 and its companions KUG 0313--006, UGC\,2628, and WISEA J031630.50--002345.1 marked on a SDSS $g$-band image with the blue contours showing the VLA BC-configuration \hi\ at column densities of 1.7, 3.5, 7.0, 14.0, 21.0, 28.0, and 35.0 $\times$ 10$^{20}$ atoms cm$^{-2}$, with the first contour at 3 $\sigma$. The small ellipse in the bottom left corner indicates the size and orientation of the VLA BC-configuration 9.5\prin\ $\times$ 8.3\prin\ FWHP synthesised beam. The large circle shows the  Effelsberg 9\prim\ FWHP  beam. \textbf{\textit{Right:}} \hi\ velocity field with contours separated by 5 \km, with velocities shown by the colour scale.}
\label{fig_3}
\end{center}
\end{figure*}

A NED search within a 20 arcmin (564 kpc) radius and $\pm$ 200 \km\ of J0315-0024 (SDSS J031559.89-002425.7 V$_{opt}$ = 6771 \km) revealed {seven} galaxy {candidates} projected within  9 arcmin (254 kpc) and a galaxy group {(LDCE 0236)} projected 13 arcmin (367 kpc) from the BCD. The nearest {galaxy candidates are} the highly optically disturbed {system of galaxies,  KUG 0313-006. This merging system contains five NED galaxy identifications} projected $\sim$ 2.5 arcmin (70 kpc) SE of J0315-0024 with optical velocities in the range of 6815 \km\ to {6893} \km. The closest galaxy in terms of velocity is the small galaxy,  WISEA J031630.50-002345.1 (V$_{opt}$ = 6784\km), but this latter is projected 7.7 arcmin (217 kpc) to the east of the BCD. Even further to the SE is the large barred spiral  UGC\,2628 (V$_{opt}$ = 6814 \km), which is projected 8.8 arcmin (248 kpc) from the BCD. {The most distant companion is WISEA J031559.90-002425.5, V$_{opt}$ = 6939 \km, which is  projected 15.5 arcmin (437 kpc) from the BCD. Both  WISEA J031559.90-002425.5 and LDCE 0236 appear unlikely to have recently  interacted with J0315-0024 based on their projected separations from the BCD.} The projected positions of the BCD and its nearest companions are shown in Fig. \ref{fig_3}. The DECaLS $g$-band image shows a  spur extending along the  eastern side {of the} galaxy, which appears to {be} an  interaction signature, possibly caused {by} a merging satellite galaxy; {see the  corresponding panel of Fig. \ref{fig_a1}.}  
\subsection{J2053+0039}
\label{sample_J2053s}
\begin{figure*}
\begin{center}
\includegraphics[ angle=0,scale=0.6] {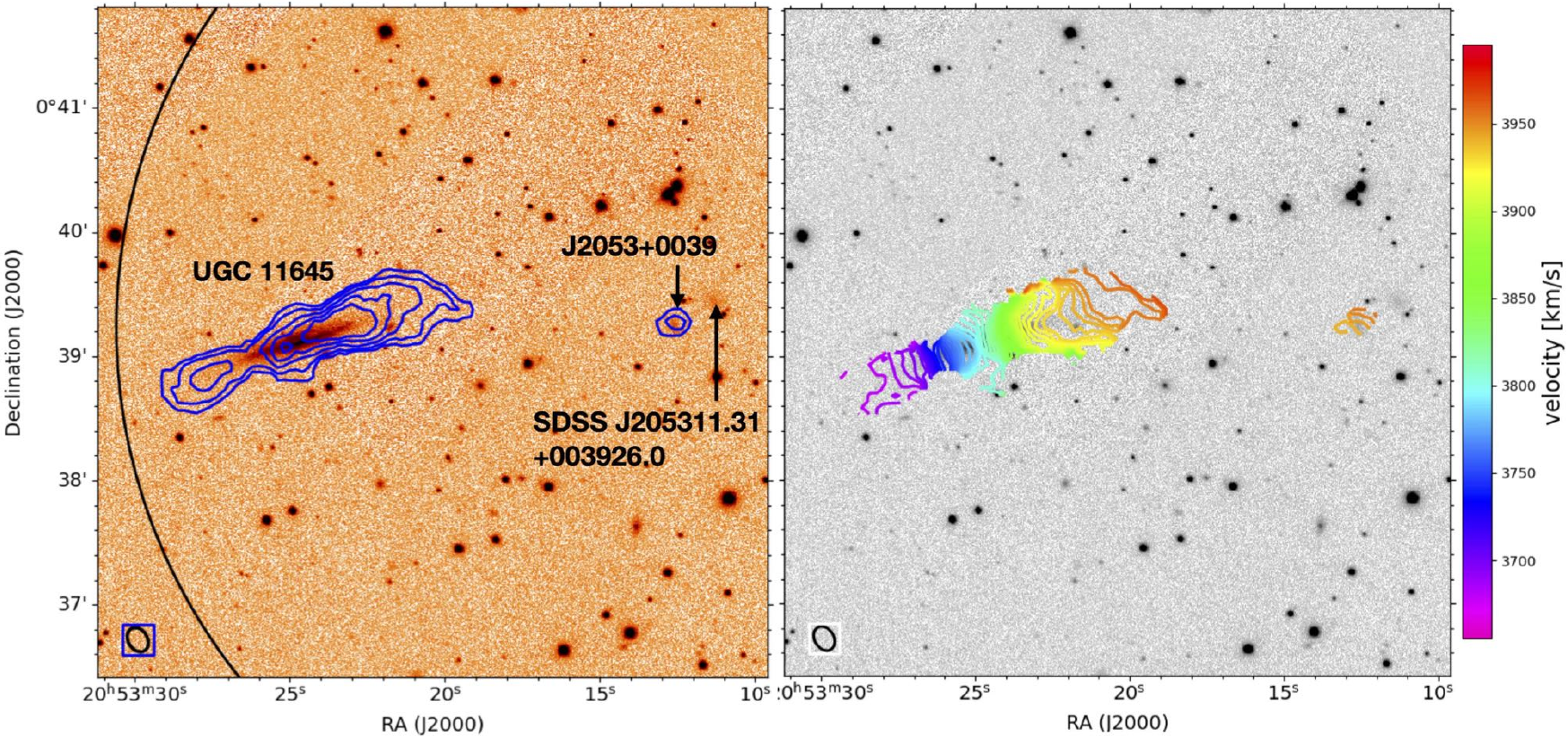}
\vspace{0.01cm}
\caption{\textbf{\textit{Left:}}  \textcolor{black}{J2053+0039 field.}
hi\ detected in the BCD target J2053+0039  and its companion UGC\,11645, shown with blue contours. The VLA BC-configuration \hi\ contours are at column densities of 2.3, 4.6, 9.0, 14.1, and 18.2  $\times$ 10$^{20}$ atoms cm$^{-2}$, with the first contour at 3 $\sigma$. The IDs of the galaxies are marked on an SDSS  $g$-band image. The small ellipse in the bottom left corner indicates the size and orientation of the VLA BC-configuration 12.4\prin\ $\times$ 9.3\prin\ FWHP synthesised beam.  The large circle shows the  Effelsberg 9\prim\ FWHP beam. \textbf{\textit{Right:}} \hi\ velocity field with contours separated by 5 \km, with velocities shown by the colour scale.}
\label{fig_4}
\end{center}
\end{figure*}

Searching NED within a 30 arcmin (538 kpc) radius and $\pm$ 200 \km\ of J2053+0039 (V$_{opt}$ = 3935$\pm$2 \km) reveals only one  object, UGC 11645, which is a small (optical diameter 18 kpc) edge-on Sbc spiral galaxy projected 3.0 arcmin (54 kpc)  east of J2053+0039 with  V$_{opt}$ = 3815 \km; see Fig. \ref{fig_4}.  The  DECaLS $g$-band image of J2053+0039 {and its DECaLS colour image (Fig. \ref{fig_a1})} suggest a cometary optical morphology with the tail oriented to the north.  There is also emission from an object $\sim$ 22 arcsec (6.8kpc) to the NW, which appears to be similar in size and surface brightness to J2053+0039. In  SDSS DR16, this second object, SDSS J205311.31+003926.0 (20 5311.31, +00 39 26.04), is classified as a galaxy with a photometric redshift of 0.119 $\pm$0.0727 ($\sim$ 33625\km). Although, based on its {DECaLS $g$-band image alone,}  J205311.31+003926.0 seems closer than that, {but}  lacks a spectroscopic redshift to confirm this. There is no evidence that it tidally interacted with J2053+0039 in the recent past.  


\begin{figure*}
\begin{center}
\includegraphics[ angle=0,scale=0.6] {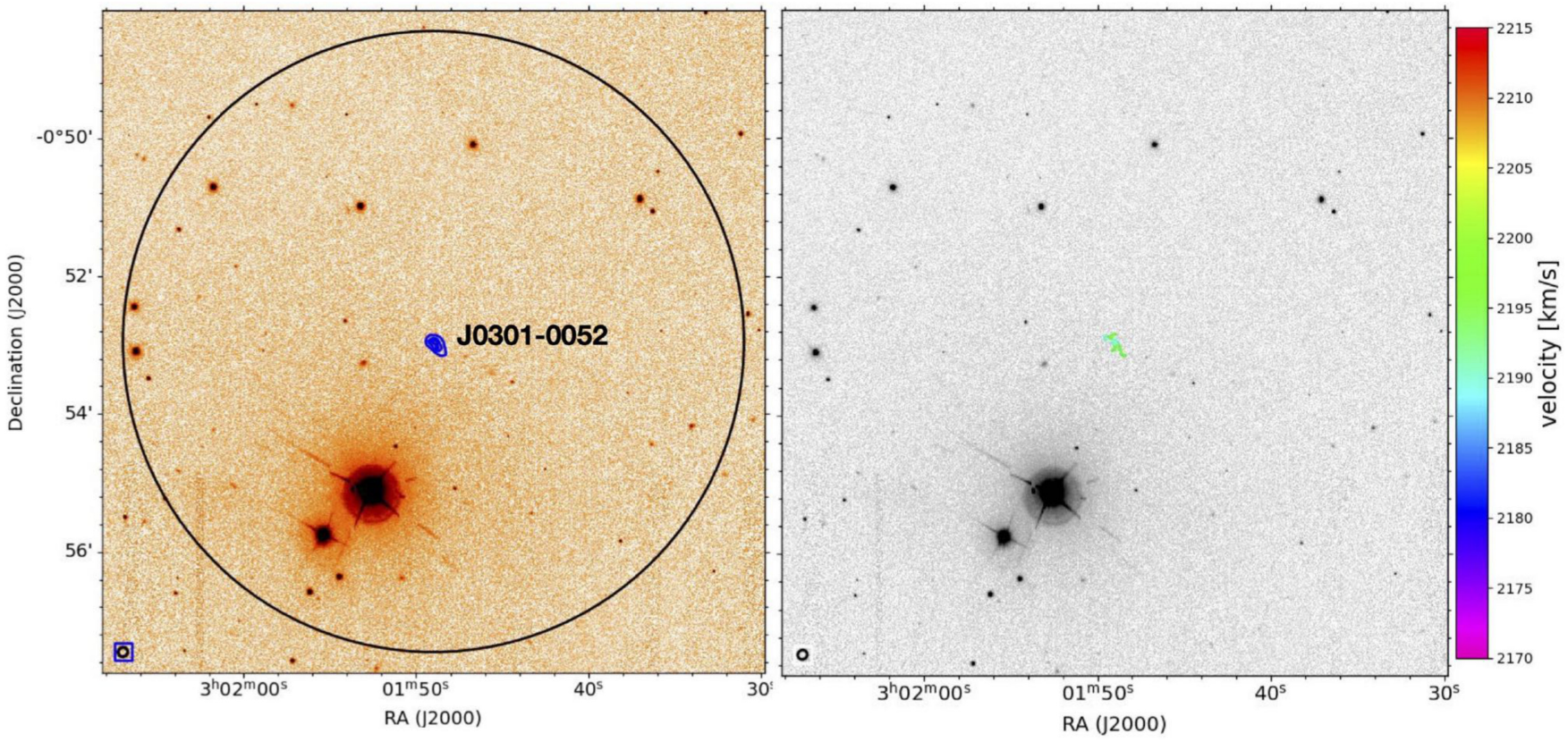}
\vspace{0.01cm}
\caption{\textbf{\textit{Left:}} \textcolor{black}{J0301--0052 field.}
\hi\ detected in the BCD target J0301--0052, marked on an SDSS $g$-band image with the blue contours showing  the VLA BC-configuration \hi\ at column densities of 2.0, 4.0, and 5.3 $\times$ 10$^{20}$ atoms cm$^{-2}$, with the first contour at 3 $\sigma$. The small ellipse in the bottom left corner indicates the size and orientation of the VLA BC-configuration 8.4\prin\ $\times$ 7.9\prin\ FWHP synthesised beam. The large circle shows the  Effelsberg 9\prim\ FWHP beam. \textbf{\textit{Right:}} \hi\ velocity field with contours separated by 5 \km, with velocities shown by the colour scale.}
\label{fig_2}
\end{center}
\end{figure*} 


\section{observations and reduction}
\label{obs} 
Each of our four BCD targets was observed with the VLA (project ID: 20A--005) for a total of 6.3 h on-source (4.7 h in B-configuration and 1.6 h in C-configuration) in L-band  (1420\,MHz). A summary of the observation from each day is set out in Table \ref{table2}. The pointing centre of each observation was at the position of  {its} BCD target.

The C-configuration observations were intended to provide surface brightness sensitivity for the diffuse \hi\ and potentially evidence for \hi\ infall and/or interactions with neighbours. The  B-configuration observations were made to detect finer but brighter details and had integration times that were three times longer than the  C-configuration to partially compensate for the lower B-configuration surface brightness sensitivity.

Each scheduling block's observation was configured to record 16 plus 2 spectral windows (spws). The 16 spws covered 64\,MHz each in 1\,MHz wide channels to record the continuum emission at full polarisation across the 1--2 GHz L-band front-end. The two additional spws were each 4\,MHz in width, split into 1024 channels of 3.90625\,kHz in width, and used for spectral line observations in dual polarisation mode.  One of the spectral line spws was set up to observe the neutral atomic hydrogen \hi\ line, while the other one was centred on the main {OH} line transitions at 1665/1667\,MHz. No OH emission was detected in our calibrated data and no further use was made of this spw. The continuum data will be analysed and published separately. In this paper, we concentrate on the \hi\ results.

The measurement set  of each scheduling block was reduced with the {\sc CASA calibration pipeline} (6.2.1.7) provided within the Common Astronomy Software Applications package {\sc (casa}; 6.2.1.7) following standard procedures \citep{mcmull07}. {We verified that the \hi\ observations of  each scheduling block reached the required quality and expected noise value. Our aim is only to obtain \hi\ imaging and no self calibration is required for our data.}  { Subsequently, we removed the continuum from each measurement set in the {\em uv--}domain using the {\sc casa} task {\sc uvcontsub}.} We then proceeded to combine the data for each target field and create the data cubes { ($\alpha,  \delta,  $velocity)} used for {our} analysis.

The observations are taken in the topocentric reference frame using the radio definition for the Doppler shift. { Because each scheduling block was observed at a different time or on a different date, this introduces  slight Doppler frequency shifts between the observations.} Ordinarily {these are corrected `on the fly' using} the {\sc casa} task {\sc tclean}  before gridding the visibilities and applying the Fourier transform. However, it was found that this led to the first channel of the resulting cube {being} corrupted, and as this channel sets the clean iteration cycle thresholds, this prevented proper `cleaning' of the channel maps. To remedy this, we used the {\sc casa} task {\sc cvel2} to convert each scheduling block from topocentric (TOPO) to the kinetic LSR (LSRK) reference frame, extracting 950 channels over a common{, Doppler-corrected} frequency range.  The next step was to use {\sc tclean} to combine and clean the five continuum-subtracted  \hi\ measurement sets for each target into a single combined (BC configuration)  image cube.  Each target's image cube has about 950 channels with a channel width of  3.90625\,kHz.   Because we were particularly interested in the faint \hi\ emission at the edges of the \hi\ discs, we produced the BC-configuration \hi\ cubes using natural weighting, as this optimises for sensitivity; although this comes at a  small cost in terms of angular resolution and synthesised beam shape definition.  The image cubes were then converted to a heliocentric reference frame (BARY). To improve the signal-to-noise ratio in the image cubes for each target, we used the CASA task {\sc specsmooth} to{ reduce the velocity resolution per channel to $\sim$
 5 \km, averaging six channels into one.{ The data cubes were corrected for the primary beam attenuation of the  VLA and were imported into the Astronomical Image Processing System ({\sc aips}; \citealt{wells85}). We then used the {\sc aips} tasks {\sc blank}  and {\sc xmom,} respectively, to identify the \hi\ signal, channel by channel, and finally collapse each blanked cube} to create moment 0, 1, and 2 maps, where moment 0 is the total or integrated \hi\ map, moment 1 is the velocity field, and moment 2 the velocity dispersion.}



\begin{table*}
\begin{minipage}{150mm}
        \caption{VLA \hi\ detections. }
        \label{table3}
        \begin{center}
        \begin{tabular}{lllllrrrrr}
                
                \hline

                Field &Galaxy ID &R.A.\footnote{All galaxy coordinates are from \citep{filho13} except NGC\,811,PGC\,7892 and UGC\,11645 which are from NED.}&
                Dec.&
                V$_{HI}$ &
                W$_{20}$&
                {W$_{50}$}&
                {M(HI)}\footnote{\mhi\ = 2.36 $\times$ 10$^{5}$D$^{2}$ $\int$S$_{\nu}$ dv \citep{giovan85}. D is the distance in Mpc and  $\int$S$_{\nu}$ dv is the integrated flux from our VLA observations, except for UGC\,11645 where we use the \hi\ flux (4.49 Jy) from \citet{masters14} because the VLA flux for that galaxy was contaminated by RFI.}&
                {M(HI)/M$_*$}\footnote{{Where M$_*$ is from \cite{filho13}.}} & 
                $g-i$\footnote{SDSS model magnitude colours.} \\
                &&[h\,m\,s]&[d\,m\,s]&[\km]&[\km]&{[\km]}&[10$^8$\msolar]&\\
                \hline
                
                J0204&J0204--1009 & 02 04 25.6&$-$10 09 36 & 1904$\pm$3 &80$\pm$5&{65$\pm$5} & {3.04}&{15.2}&0.15 \\
                J0204&NGC\,811 & 02 04 34.8 & $-$10 06 30.52 &1924$\pm$6 & 175$\pm$12& {150$\pm$12}& {35.93} \\
                J0204&PGC\,7892& 02 04 12.4 &  $-$10 04 42.60& 1902$\pm$4 &95$\pm$7&{75$\pm$7}& {9.34} \\
                J0204&WISEA\footnote{{Full ID, WISEA\,J020448.18-095859.0}}& 02 04 48 &  $-$09 59 03& 1879$\pm$5 &75$\pm$10&{50$\pm$10}& {4.55}\\
                J0301&J0301-0052&03 01 49.0&$-$00 52 57 & 2193$\pm$3 &30$\pm$6&{30$\pm$6}& 0.06&{0.2}&-0.22 \\    
                J0315&J0315-0024 & 03 15 59.9& $-$00 24 26& 6774$\pm$2 & 83$\pm$4& {78$\pm$4}& { 5.99}&{12.0}&-0.07 \\
                J0315&KUG\,0313-006& 03 16 06.10& $-$00 26 21& 6797$\pm$10 & 202$\pm$20& {176$\pm$20}&{57.91} \\
                J0315&UGC\,2628& 03 16 31.8& $-$00 28 05& 6810$\pm$4& 404$\pm$9& {388$\pm$9}& {193.47} \\              
                J2053&J2053+0039 &20 53 12.6&$+$00 39 14 &3944$\pm$4  & 56$\pm$8& {36$\pm$8}& {0.23}&{1.2} &0.28 \\
                J2053&UGC\,11645 &20 53 24.6 &$+$00 39 07.05 & 3815$\pm$4 & 306$\pm$9& {299$\pm$9}& 149.59& \\
                        
                \hline
\end{tabular}
\end{center}
\end{minipage}
\end{table*}

\section{Observational results}
\label{results}
The \hi\ results for the combined VLA BC configuration image cubes for each target field are summarised below, with the V$_{HI}$, W$_{20}$, {W$_{50}$}, \mhi,\ and \mhi/M$_*$ reported in Table \ref{table3}. Spectra extracted from the VLA BC-configuration cubes for the four BCD galaxies  are presented in Fig. \ref{fig_6}.
 
\subsection{J0204--1009}
\label{204_result}
\begin{figure*}

\includegraphics[ angle=0,scale=0.30,width=0.80\textwidth] {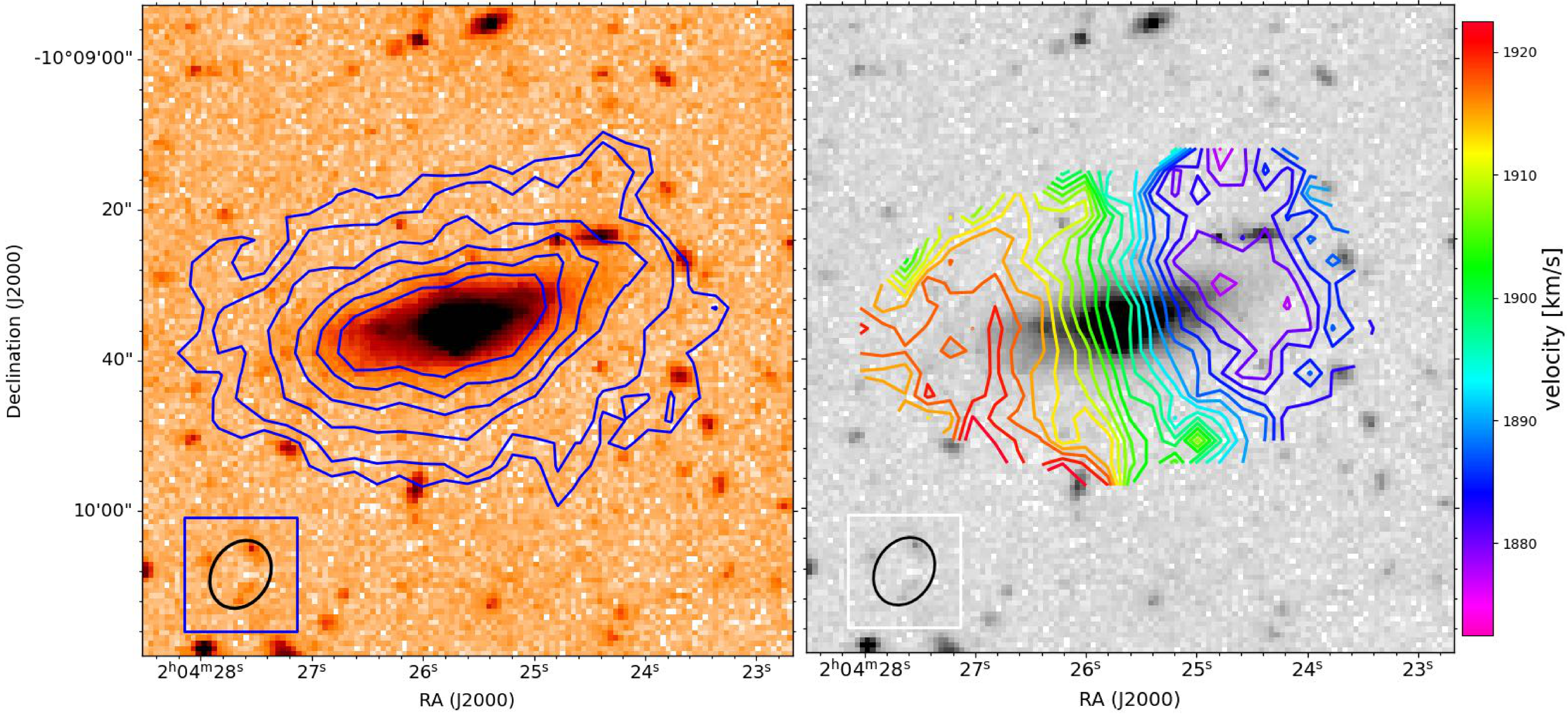}
\caption{\textcolor{black}{J0204--1009. }\textbf{\textit{Left:}} 
BCD J0204--1009 CFHT $g$-band image  with VLA BC-configuration moment 0   \hi\ contours overlaid. The VLA BC-configuration \hi\ contours are at column densities of \textcolor{black}{2.7}, 4.1, 8.2, 12.3, 16.5,  and 20.6 $\times$ 10$^{20}$ atoms cm$^{-2}$, with the first contour at 3 $\sigma$. \textbf{\textit{Right:}} \hi\ velocity field with contours separated by \textcolor{black}{2.5} \km. The ellipse in the bottom left corner of each panel indicates the size and orientation of the VLA BC-configuration 9.5\prin\ $\times$ 7.6\prin\ FWHP synthesised beam. }
\label{fig_5_204}
\end{figure*}

\begin{figure*}
\includegraphics[ angle=0,scale=0.62,width=0.80\textwidth] {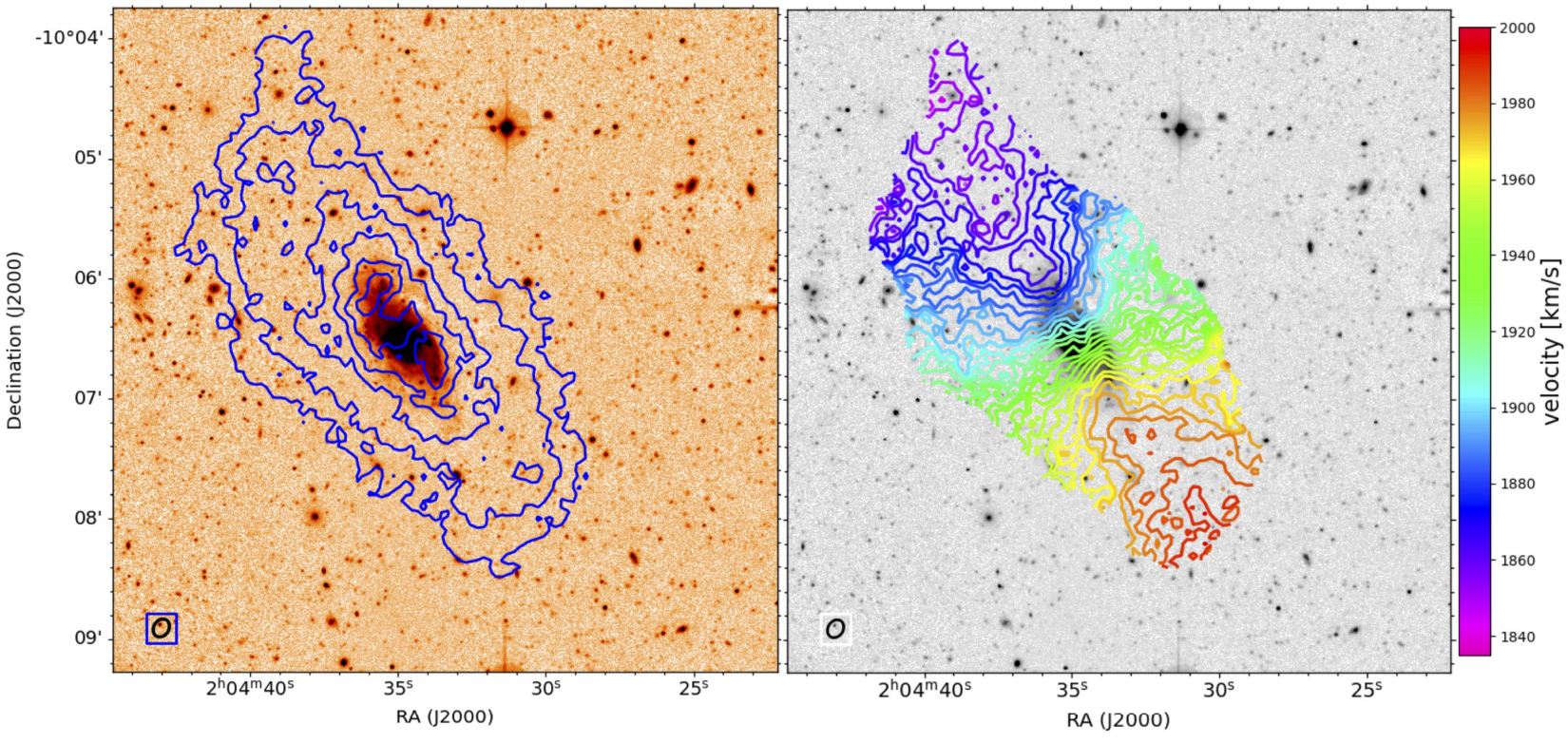}
\vspace{0.01cm}
\caption{NGC\,811  in the BCD J0204--1009 field. \textbf{\textit{Left:}} Moment 0 \hi\ contours on the CFHT $g$-band image. The VLA BC-configuration \hi\ contours are at column densities of \textcolor{black}{3.1, 7.6, 15.2, 22.9, 30.5, and 38.1} $\times$ 10$^{20}$ atoms cm$^{-2}$, with the first contour at 3 $\sigma$.  \textbf{\textit{Right:}} Velocity field contours, with a contour separation of \textcolor{black}{2.5} \km. The ellipse in the bottom left corner of each panel indicates the size and orientation of the VLA BC-configuration 9.5\prin\ $\times$ 7.6\prin\ FWHP synthesised beam.}
\label{fig_811}
\end{figure*}

\begin{figure*}
\includegraphics[ angle=0,scale=0.74,width=0.80\textwidth] {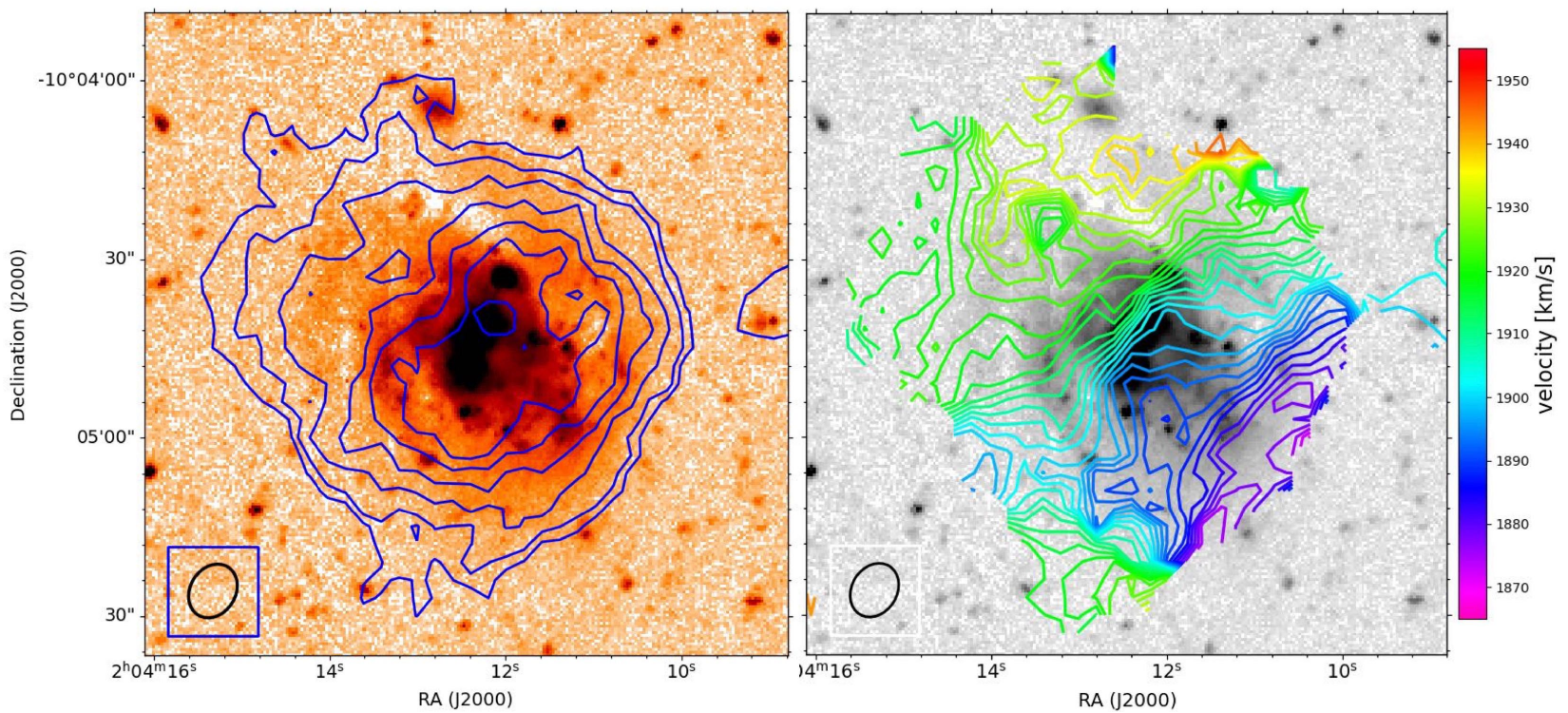}
\vspace{0.01cm}
\caption{PGC\,7892  in the BCD J0204--1009 field. \textbf{\textit{Left:}} Moment 0 \hi\ contours on a CFHT $g$-band image. The VLA BC-configuration \hi\ contours are at column densities of 2.1, 4.1, 8.2, 12.3, 16.5, 20.6, and 41.1 $\times$ 10$^{20}$ atoms cm$^{-2}$, with the first contour at 3 $\sigma$. \textbf{\textit{Right:}} Velocity field contours, with a contour separation of  \textcolor{black}{2.5} \km. The ellipse in the bottom-left corner of each panel indicates the size and orientation of the VLA BC-configuration 9.5\prin\ $\times$ 7.6\prin\ FWHP synthesised beam.}
\label{fig_7892}
\end{figure*}

\begin{figure*}
\includegraphics[ angle=0,scale=0.28,width=0.80\textwidth] {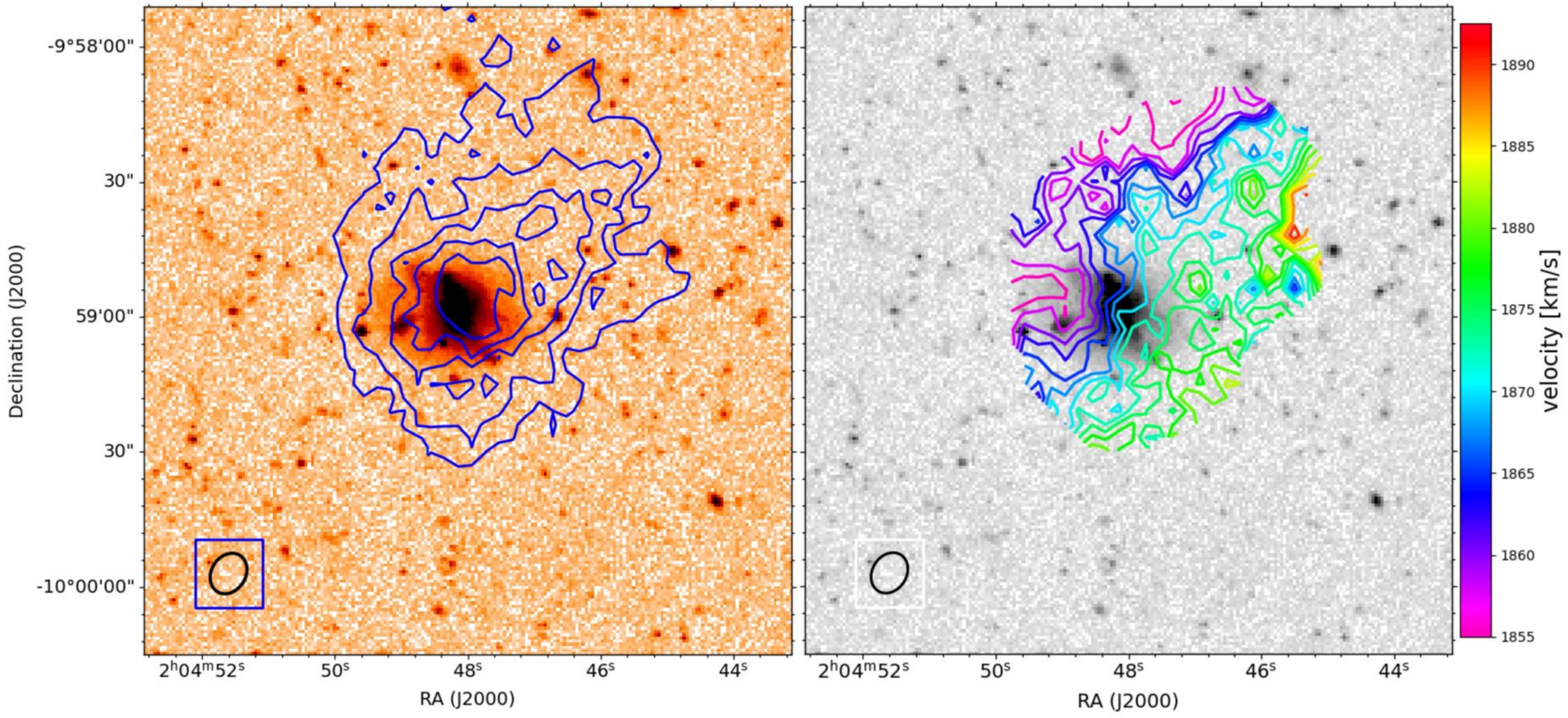}
\vspace{0.01cm}
\caption{WISEA\,J020448.18-095859.0  in the BCD J0204--1009 field. \textbf{\textit{Left:}} Moment 0 \hi\ contours on CFHT $g$-band image. The VLA BC-configuration \hi\ contours are at column densities of \textcolor{black}{4.4, 8.2, 12.3, 16.5 and 20.6}  $\times$ 10$^{20}$ atoms cm$^{-2}$, with the first contour at 3 $\sigma$. \textbf{\textit{Right:}} velocity field contours, with contour separation of  \textcolor{black}{2.5} \km. The ellipse in the bottom-left corner of each panel indicates the size and orientation of the VLA BC-configuration 9.5\prin\ $\times$ 7.6\prin\ FWHP synthesised beam.}
\label{fig_wisea}
\end{figure*}

Our wide-field VLA BC-configuration integrated \hi\ (moment 0) map and velocity field, overlaid as contours on a CFHT $g$-band image, as shown in Fig. \ref{fig_1}, show \hi\ detections for our BCD target J0204--1009 (V$_{HI}$ = { 1904$\pm$3} \km) and two more massive companion galaxies.  These latter are
NGC\,811, at a velocity of V$_{HI}$ = {1924$\pm$6} \km, projected 3.8 arcmin (30 kpc) to the NE of J0204--1009, and PGC\,7892, at V$_{HI}$ = {1902$\pm$4} \km, projected 5.9 arcmin (47 kpc) to the NW of J0204--1009. A dwarf companion,  WISEA\,J020448.18-095859.0 (PGC\,983900), V$_{HI}$ = {1879$\pm$5} \km,  12 arcmin (a projected distance of 95 kpc) to the NE of J0204--1009 was also detected in \hi\ within the VLA 32\prim\  FWHP primary beam. All four  \hi\ detections are part of the PGC\,7998 group \citep{kourkchi17}.  Figures \ref{fig_5_204}, \ref{fig_811}, \ref{fig_7892}, and \ref{fig_wisea} show zoomed-in views of the moment 0 and velocity fields for each \hi\ detection.   

The optical and \hi\ morphologies and in particular the \hi\ velocity fields for all four galaxies are consistent with recent interactions within the group  {and} within the \hi\ relaxation timescale, that is, within $\sim$ 0.7 Gyr \citep{holwerda11}. Whereas the \hi\ morphologies suggest all four detections have been subject to interactions, the velocity fields suggest an interaction with WISEA\,J020448.18-095859.0 has perturbed the NE part of the \hi\ disc of NGC\,811. On the other hand, the velocity range of J0204--1009 is consistent with an interaction with PGC\,7892. Figure \ref{fig_5_204}  shows that the \hi\ morphology of J0204--1009 is symmetric to  first order. {Overall,} its  velocity field (right panel)  presents {an \hi\ disc in regular rotation}. However, the change in {the PA of the iso-velocity contours} at the western edge of the disc and closed iso-velocity contours at velocities \textcolor{black}{at} around 1870 \km\ to 18\textcolor{black}{8}0 \km\  is an indication of a significant warp, possibly caused {by} a recent interaction with PGC\,7892. The SE
morphology of PGC\,7892  and its truncated velocity field at $\sim$ 1870 \km\ to 18\textcolor{black}{8}0 \km\ are consistent {with a} recent flyby interaction with J0204--1009.  In both potential interaction pairs, the outer disc of the larger galaxy shows a larger perturbation than that visible in the smaller galaxies, which is perhaps related to a shorter \hi\ relaxation timescale for the less massive galaxies.  

\subsection{J0301-0052}
\label{301_result}
\begin{figure*}
\includegraphics[ angle=0,scale=0.31,width=0.80\textwidth] {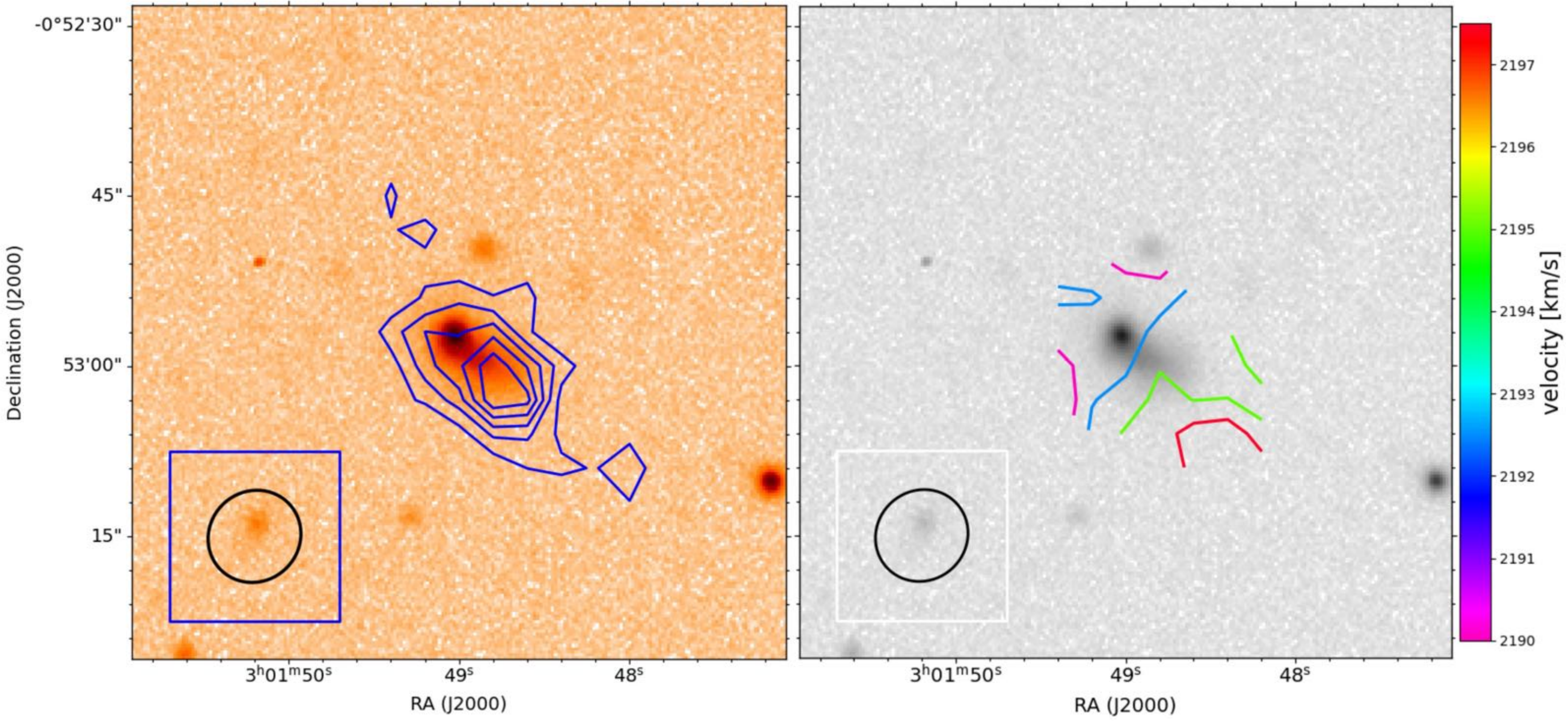}
\vspace{0.01cm}
\caption{\textcolor{black}{J0301-0052. }  \textbf{\textit{Left:}} 
BCD J0301-0052 DECaLS $g$-band image  with VLA BC-configuration moment 0   \hi\ contours overlaid. The VLA BC-configuration \hi\ contours are at column densities of \textcolor{black}{3.2, 4.1, 5.0, 5.8  and 6.6} $\times$ 10$^{20}$ atoms cm$^{-2}$, with the first contour at 3 $\sigma$.   \textbf{\textit{Right:}} \hi\ velocity field with contours separated by \textcolor{black}{2.5} \km. The ellipse in the bottom-left corner of each panel indicates the size and orientation of the VLA BC-configuration 8.4\prin\ $\times$ 7.9\prin\ FWHP synthesised beam.}
\label{fig_5_301}
\end{figure*}

J0301-0052 was the sole \hi\ detection in the VLA BC-configuration {field of view} (FOV; Fig. \ref{fig_2}). The \hi\ detection at the position of the BCD J0301-0052 (SDSS J030149.02-005257.3) has a velocity of 2193$\pm$3 \km. This is in close agreement with the V$_{opt}$ = 2194 \km\ from NED but is 85 \km\ higher than that reported in \cite{filho13}. Additionally, the VLA value of {W$_{50}$} of 30$\pm$6 \km\ is lower than W$_{50}$ = 110 \km\ reported for this object by \cite{filho13};
 see also the \hi\ spectrum shown in  Fig. \ref{fig_6}.  A zoom-onto the \hi\ moment 0 map (Fig. \ref{fig_5_301}) reveals a \hi\ counterpart \textcolor{black}{which like} the optical galaxy has a cometary morphology,
although the \hi\ tail is more extended and {the \hi\ column density maximum is projected $\sim$ 6 arcsec ($\sim$ 1 kpc) SW of the optical centre, near the end of the optical tail. While the \hi\ -- optical maxima offset is smaller than the VLA synthesised beam,  the signal-to-noise ratio of the \hi\ maxima makes the offset convincing.  Although the velocity field lacks resolution, the velocity contours near the column density maximum  appear {to be} disturbed.}

\begin{figure*}
\includegraphics[ angle=0,scale=0.64,width=0.80\textwidth] {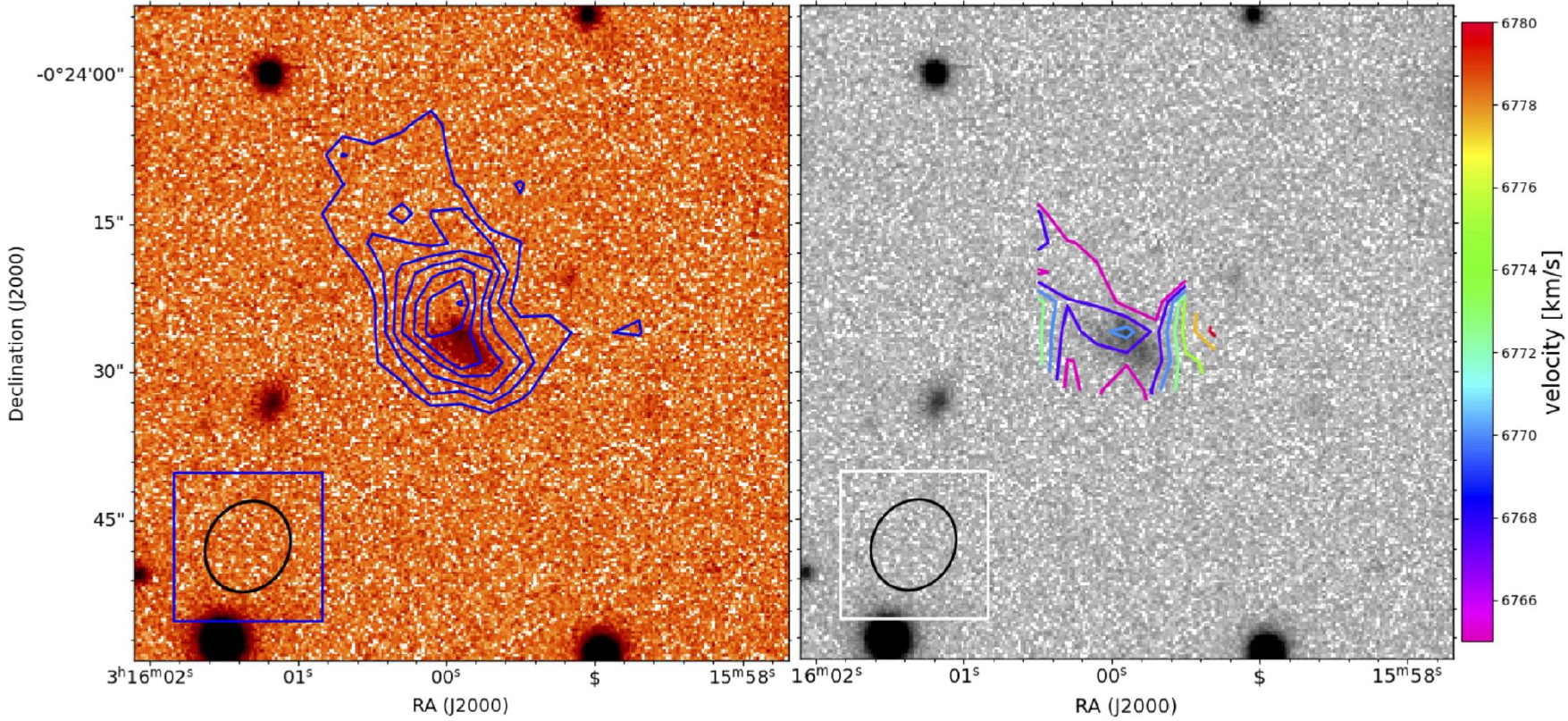}
\vspace{0.01cm}
\caption{\textcolor{black}{J0315-0024. }\textbf{\textit{Left:}} 
BCD J0315-0024 DECaLS $g$-band image  with VLA BC-configuration moment 0   \hi\ contours overlaid. The VLA BC-configuration \hi\ contours are at column densities of \textcolor{black}{2.0, 2.8, 3.5, 4.2, 4.9, 5.6,  and 6.2} $\times$ 10$^{20}$ atoms cm$^{-2}$, with the first contour at 3 $\sigma$. \textbf{\textit{Right:}} \hi\ velocity field with contours separated by \textcolor{black}{2.5} \km. The ellipse in the bottom-left corner  of each panel indicates the size and orientation of the VLA BC-configuration 9.5\prin\ $\times$ 8.3\prin\ FWHP synthesised beam.} 
\label{fig_5_315}
\end{figure*} 


\begin{figure*}
\includegraphics[ angle=0,scale=0.28,width=0.80\textwidth] {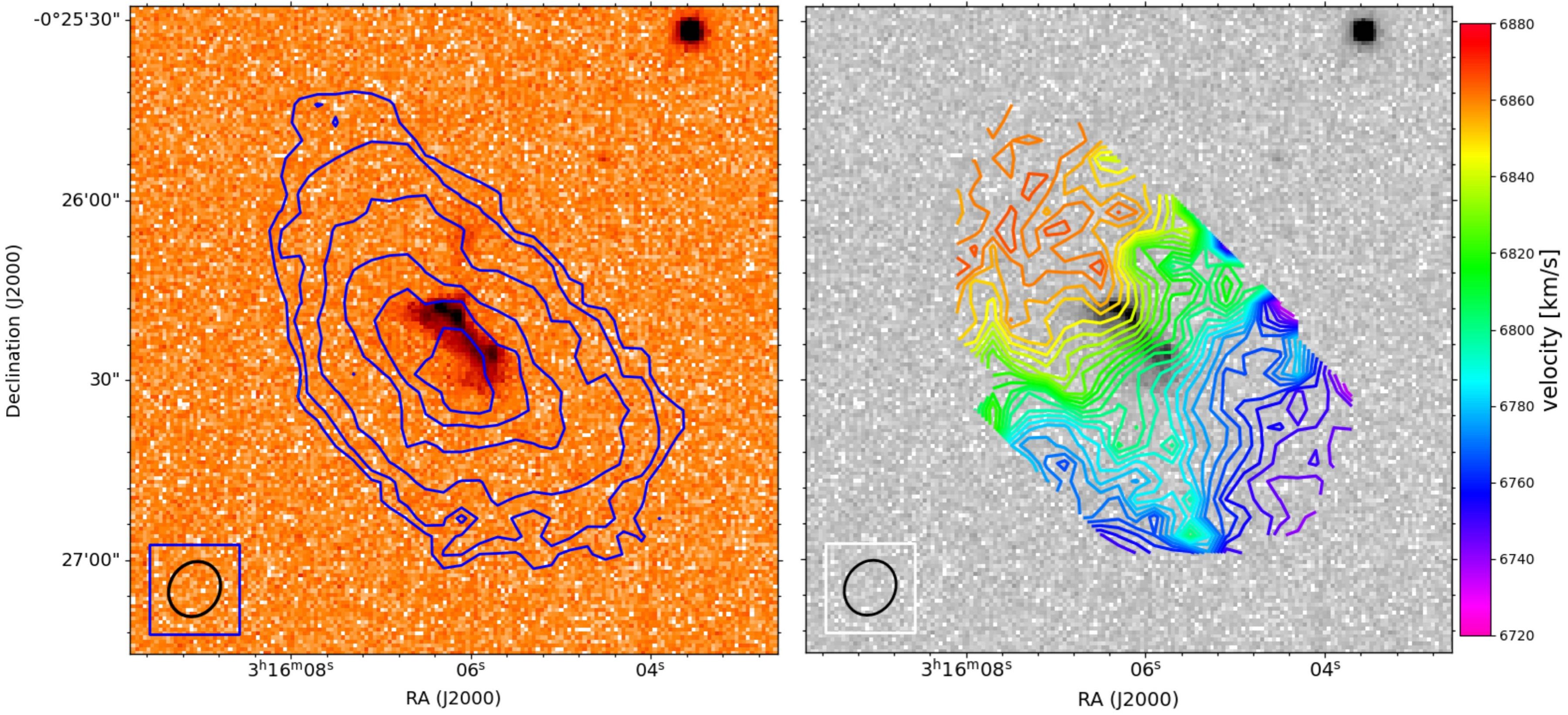}
\vspace{0.01cm}
\caption{KUG\,313-006  in the J0315--0024 field. \textbf{Left:} \hi\ contours overlayed on an SDSS $g$-band image. The  VLA BC-configuration \hi\ at column densities of \textcolor{black}{2.1, 3.5, 7.0 14.0, 20.9, and 27.9} $\times$ 10$^{20}$ atoms cm$^{-2}$, with the first contour at 3 $\sigma$. The small ellipse in the bottom-left corner indicates the size and orientation of the VLA BC-configuration 9.5\prin\ $\times$ 8.3\prin\ FWHP synthesised beam. \textbf{Right: }Velocity field contours, with a contour separation of  \textcolor{black}{2.5} \km}
\label{fig_10a}
\end{figure*} 


\begin{figure*}
\includegraphics[ angle=0,scale=0.63,width=0.80\textwidth] {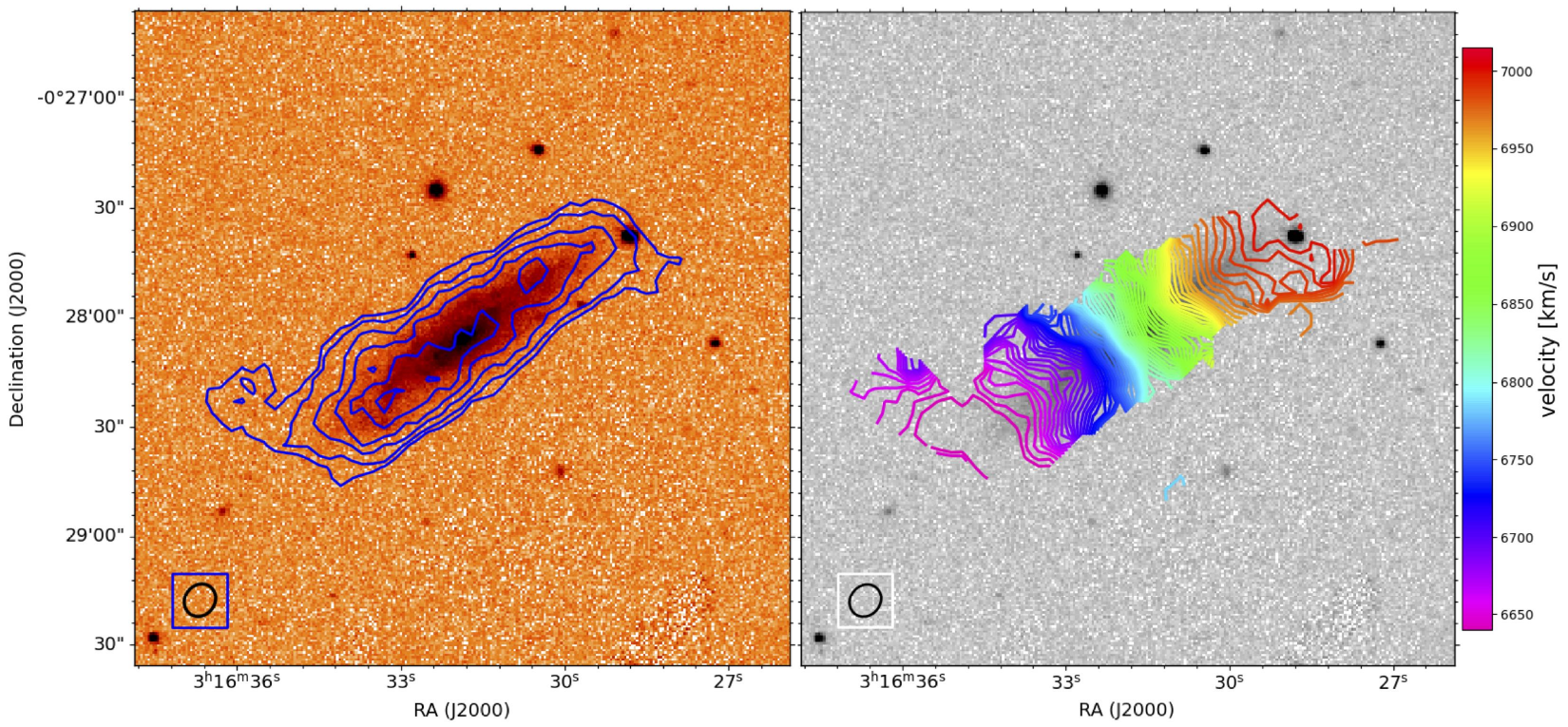}
\vspace{0.01cm}
\caption{ UGC\,2628  in the J0315--0024 field. \textbf{Left:} \hi\ contours overlayed on an SDSS $g$-band image. The  VLA BC-configuration \hi\ at column densities of 3.5, 7.0, 14.0, 21.0 28.0, and 35.0 $\times$ 10$^{20}$ atoms cm$^{-2}$, with the first contour at 3 $\sigma$. \textbf{Right:} Velocity field contours, with a contour separation of  5 \km. The ellipse in the bottom-left corner of each panel indicates the size and orientation of the VLA BC-configuration 9.5\prin\ $\times$ 8.3\prin\ FWHP synthesised beam.}
\label{fig_11}
\end{figure*} 


\begin{figure*}
\includegraphics[ angle=0,scale=0.3,width=0.80\textwidth] {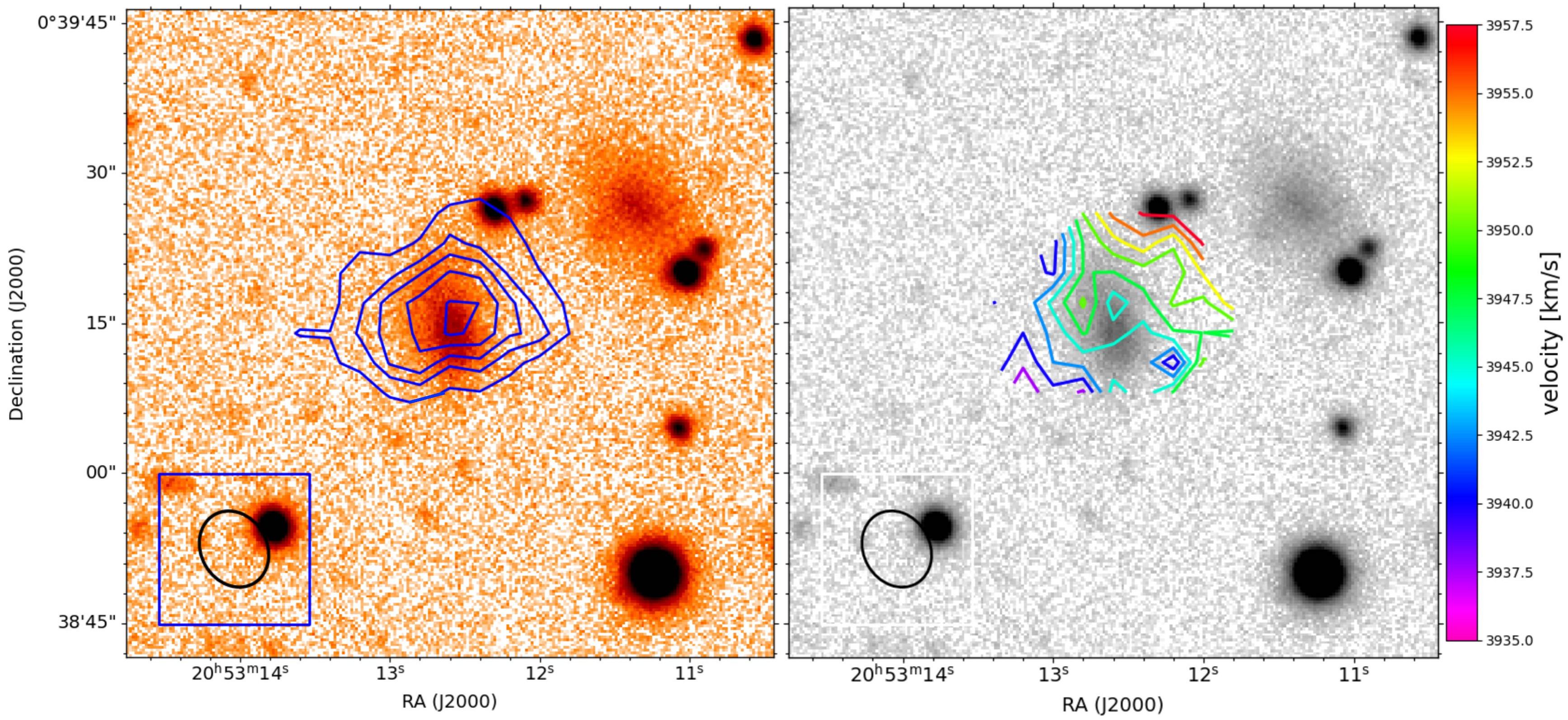}
\vspace{0.01cm}
\caption{\textcolor{black}{J2053+0039. } \textbf{\textit{Left:}} 
BCD J2053+0039 DECaALs $g$-band image  with VLA BC-configuration moment 0   \hi\ contours overlaid. The VLA BC-configuration \hi\ contours are at column densities of 1.1, 3.2, 5.3, 7.4, and 9.5 $\times$ 10$^{20}$ atoms cm$^{-2}$, with the first contour at 3 $\sigma$. \textbf{\textit{Right:}} \hi\ velocity field with contours separated by \textcolor{black}{2.5} \km. The ellipse in the bottom-left corner of each panel indicates the size and orientation of the VLA BC-configuration 7.9\prin\ $\times$ 6.6\prin\ FWHP synthesised beam.} 
\label{fig_5_2053}
\end{figure*}  

\subsection{J0315-0024}
\label{315_result}
Figure \ref{fig_3} (left panel) shows the \hi\  moment 0 contours for the VLA BC-configuration field. \hi\ was detected in J0315-0024 (V$_{HI}$ = {6774$\pm$2} \km) and in two neighbouring {objects}, KUG 0313-006  (V$_{HI}$ = {6797$\pm$10} \km) projected $\sim$ {2.6} arcmin ({73} kpc) SE of  J0315-0024 (Fig. \ref{fig_10a}) and UGC\,2628 (V$_{HI}$ = 6810$\pm$4 \km) projected $\sim$ {8.8} arcmin (248 kpc) SE of J0315-0024 (Fig. \ref{fig_11}). The  { five galaxy candidates from NED} within KUG 0313-006  form a merging system. The \hi\ moment 0 contours for the zoom-in of the BCD (Fig. \ref{fig_5_315}) show the \hi\ column density maximum offset {to the} {northern edge} of the optical galaxy{, with a one-sided \hi\  tail extending $\sim$ 24 arcsec (11 kpc) to the NE of the optical centre. This \hi\ tail appears to be an extended counterpart to the  optical spur to the east of the optical galaxy.} {The optical spur and the \hi\ tail have the same  orientation.}  The  \hi\ spectrum for this BCD is shown in Fig. \ref{fig_6}. Figure \ref{fig_3} also shows overlays of the moment 0 contours on an optical image of KUG 0313-006  and UGC\,2628. An interesting point to note from these figures is the clear displacement of the \hi\ to the {SE}  of the optical discs and the disturbed \hi\ velocity field in the KUG\,0313-006 merging system. {Also, the   morphological and kinematic warping seen at both edges of the UGC\,2628 \hi\ disc are likely signatures of  a recent ($\lesssim$ 0.7 Gyr) interaction with another group member or satellite galaxy \citep[see][]{scott14,Sanchez23}. }

\subsection{J2053+0039}
\label{2053_result}
BCD J2053+0039 is detected in {\hi\ within} the VLA BC-configuration FOV with V$_{HI}$ = 3944$\pm$4 \km,  as is a more massive edge-on companion, UGC\,11645 with V$_{HI}$ = 3815$\pm$4 \km\  (see Fig. \ref{fig_4}). Figure \ref{fig_5_2053} shows {a} zoom onto the \hi\ moment 0 contours for J2053+0039.  The contours indicate the highest column density \hi\ is slightly offset to the north of the optical BCD.    \hi\ was not detected in SDSS J205311.31+003926.0 (referred to in Sect. \ref{sample_J2053s})  projected to the NW of J2053+0039 in the velocity range of the VLA  of 3526 \km\ to 4278 \km. Given the {slightly redder colour of  SDSS J205311.31+003926.0 compared to J2053+0039 (see Fig. \ref{fig_a1}), which is consistent with its photometric redshift of 0.119 and its \hi\ non-detection,  as well as the}  lack of an optical interaction signature, it is reasonable to assume SDSS J205311.31+003926.0 lies in the background of J2053+0039. The spectrum for the BCD is shown in Fig. \ref{fig_6} (bottom {right)}.

UGC\,11645  has a V$_{opt}$ = 3815\km\ and is projected 3.0 arcmin (54 kpc) to the east of the BCD. Unfortunately, several visibilities had to be flagged  because of \textcolor{black}{radio frequency interference (RFI)} in some of the channels that contained \hi\ associated with  UGC\,11645.
This flagging occurred in a few channels around a velocity of 3775 \km\ and was visible as a small group of channels with lower flux in the spectrum. We tried to mitigate the impact of the flagging by applying a taper to the impacted channels and decreasing the resolution of the remaining channels in the cube. However, this still leaves a likely modest \hi\ flux loss in the lower velocity channels in UGC\,11645.  We emphasise that the RFI problem that impacts the \hi\ map in the eastern side of UGC\,11645 (at a velocity of around 3775 \km) does not affect the \hi\ velocity range in which J2053+0039 was detected. The \hi\ velocity of J2053+0039 coincides with velocities at the western edge of the \hi\ disc of UGC\,11645 (see Fig. \ref{fig_4}). This and the disturbed \hi\ kinematics on the western side of UGC\,11645 in the same velocity range of the BCD support a recent interaction scenario.

\section{Discussion}
\label{discussion}

\subsection{Comparison with Effelsberg 100m single-dish results}
\label{compare_effelesburg}
Figure \ref{fig_1} shows that J0204--1009 and NGC\,811 are projected within the 9 arcmin FWHP Effelsberg beam and that PGC\,7892 is projected just beyond the FWHP Effelsberg beam. V$_{HI}$ = {1904$\pm$4 \km} for the BCD is 2 \km\ higher than its V$_{opt}$ from NED and {2 \km\ lower} than that reported in \cite{filho13}.{ Therefore, all three velocities agree within the uncertainties. However, the VLA } {$_{50}$} = {65$\pm$5} \km\ for J0204--1009 is lower than the W$_{50}$ = 112 \km\ from Effelsberg \citep{filho13}. Adding the VLA spectra of J0204--1009 and NGC\,811 (removing emission below 1880 \km) produces a spectrum with a W$_{50}$ = 120 \km\ and an \hi\ profile consistent with that reported from {Effelsberg}.  This strongly suggests that the most likely explanation for differences between \hi\ properties derived from the VLA and the Effelsberg 100m  observations for  J0204-1009 is contamination by \hi\ emission from NGC\,811 in the Effelsberg 9\prim\ FWHP beam (see Fig. \ref{fig_1}). 

Comparing the VLA  \hi\ masses and spectra with those {reported} in \cite{filho13}, we find that in the cases of J0204--1009,  J0315-0052, and J2053+0039, {the  {Effelsberg}  \hi\ mass attributed to their respective BCDs is likely to be contaminated to
varying extents by} \hi\ emission from larger neighbouring \hi-rich galaxies within the Effelsberg beam (large circles in Figs. \ref{fig_1}, \ref{fig_3}, and \ref{fig_4}). {However,} only for J0301--0052 can a clean comparison between the \hi\ measurement from the VLA and  Effelsberg be made because of the absence of \hi\ companions within the Effelsberg beam. 
In that case, the Effelsberg \hi\ flux reported in \cite{filho13} was 2.1 Jy \km\ compared to the {0.026} Jy \km\ from the VLA, implying the VLA recovered only $\sim$ 3\% of the Effelsberg \hi\ flux. However, as noted in Sect. \ref{301_result}, the {Effelsberg} V$_{HI}$ of 2108 \km\  falls  below the V$_{opt}$ (2194 \km) and VLA V$_{HI}$ 2193 \km\ , and the range of velocities implied
by the Effelsberg W$_{50}$ (110 \km)  reported in \cite{filho13} is outside the VLA W$_{20}$ range
.  This led us to conclude that the {Effelsberg} detection was likely not real and therefore a valid comparison with our \hi\ detection cannot be made for J0301--0052. {We tried to confirm this conclusion by checking the \hi\ Parkes All Sky Survey \citep[HIPASS;][]{barnes01}  spectrum at the position of the BCD, but unfortunately because of the high rms in the spectrum, no \hi\ was detected within $\pm$ 500 \km\ of the V$_{opt}$} of this  BCD.  

In summary, confusion of \hi\ sources within the large {Effelsberg} beam and a likely spurious detection prevent us from making a direct \hi\ comparison with the \cite{filho13} \hi\ results. Additionally, the  selection criteria we used for our \hi\ single-dish detection inadvertently bias our sample to favour BCDs that are members of groups.

\subsection{BCD \hi\ properties}

Based on the VLA observations, the \hi\ masses of our BCDs are in the range of 6  $\times$ 10$^6$ \msolar\ to 6 $\times$ 10$^8$ \msolar\ (Table \ref{table3}) with  W$_{20}$ between 30 \km\ and 83 \km. The BCDs, except J0204--1009,  present irregular \hi\ morphologies, which can be interpreted as interaction signatures with varying degrees of certainty. However, the spatial resolution for the   BCDs other than J0204--1009 is quite limited. 

In the case of J0204--1009, the strongest evidence for a recent interaction comes from the warp at the western edge of the  \hi\ velocity field rather than its \hi\ morphology (Sect. \ref{204_result}). Moreover, the two maxima in the J0204--1009 \hi\ spectrum (Fig. \ref{fig_6}) are not at the extreme edges of the velocity range, as is expected for an unperturbed high-inclination ($\sim$ 80\degree) rotating disc, indicating that the outer disc has recently been disturbed \citep[see][]{scott22}. The M$_*$ = 0.2 $\times$ 10$^8$
\msolar\ and \mhi\ = 3.04 $\times$ 10$^8$ \msolar\ of J0204--1009 give an \hi\ gas-mass-to-stellar-mass ratio of about 15, indicating an extremely gas-rich BCD.

\begin{figure*}
\includegraphics[ angle=0,scale=0.3,width=0.80\textwidth] {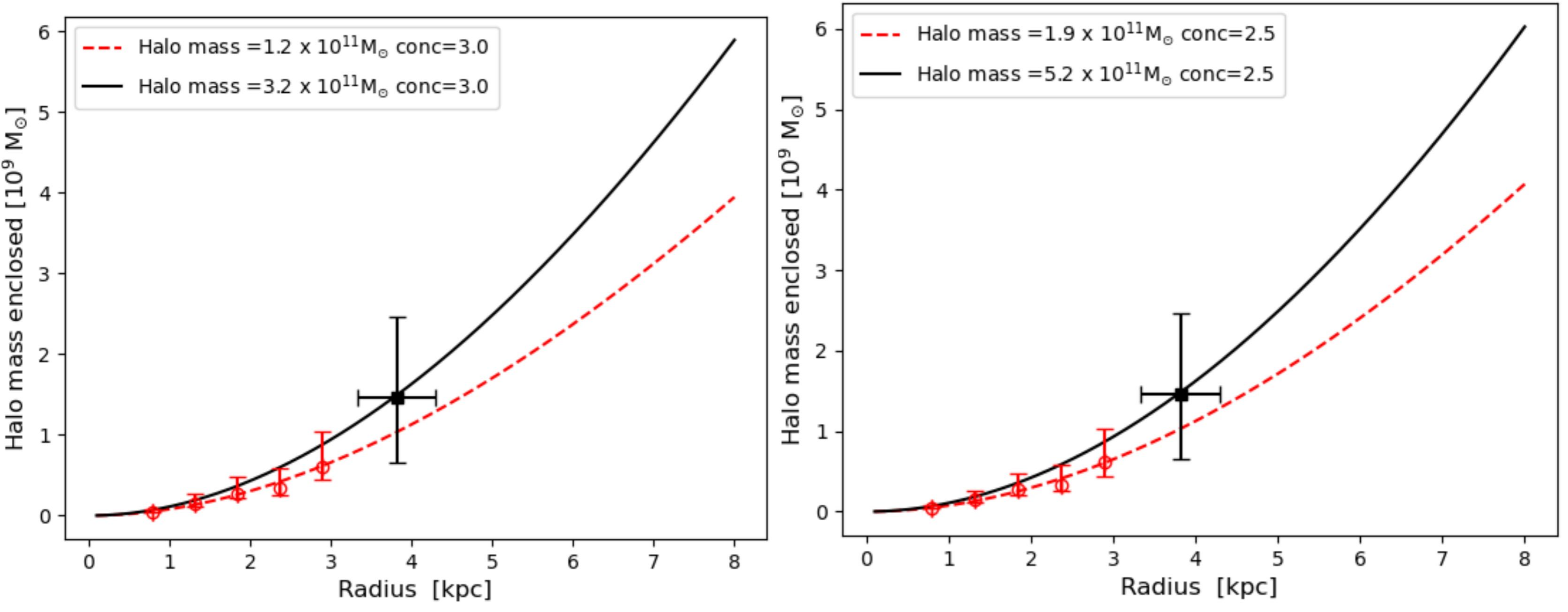}
\vspace{0.01cm}
\caption{ $M_{\mathrm dyn}$ interior to each of the {five} rings from the $^{3D}${BAROLO} fit to the \hi\ velocity field (red open circles) of J0204--1009 and  $M_{\mathrm dyn}$ derived from the $W_{20}$ value of  its integrated spectrum and \hi\ disc diameter (black square). Also shown are NFW model cumulative masses for two sets of NFW DM haloes, one set for an NFW concentration index of {3.0} (left panel) and the other for an index of {2.5} (right panel).}
\label{fig_11a} 
\end{figure*}

We used $^{3D}${BAROLO} \citep{diTeodoro2015} to fit a \textcolor{black}{five}-ring model to {the} J0204--1009 \hi\ cube with the best-fit model giving an inclination of  $\sim$ 80\degree\ and inclination-corrected maximum rotation velocity, $V_{\mathrm{rot}}$, of 30.1 \km. {We note there is a significant  uncertainty associated with the inclination derived from the  $^{3D}${BAROLO} fit, which is probably of the order of 5\degree.}  We then used the same method as detailed  in \cite{sengupta2019} to fit Navarro, Frenk, White \citep[NFW;][]{Navarro96} model dark matter \textcolor{black}{(DM)} halos to the J0204--1009 $^{3D}${BAROLO} rotation curve. Figure \ref{fig_11a} shows the fit to the $^{3D}${BAROLO}  model rotation curve. The dynamical mass, $M_\mathrm{dyn}$, was also estimated from the W$_{20}$ based on the integrated spectrum and the extent of the \hi\ disc and its fit to the NFW model. Our $M_\mathrm{dyn}$ estimate is marked in the figure with a black square. With only \textcolor{black}{five}  beams across the \hi\ disc, the NFW concentration index is poorly constrained. Dwarf galaxies are expected to have {concentration}  indices of $<$ 3. The figure shows the NFW model fits {for both the rotation curve and our estimated M$_{dyn}$}  for concentration indices 3.0 (left panel) and 2.5 (right panel). These models indicate a \textcolor{black}{DM} halo mass in the range of {1.2 $\times$ 10$^{11}$ to 5.2 $\times$ 10$^{11}$}  \msolar. The M$_*$ +  \mhi\ =  {3.24} $\times$ 10$^8$ \msolar\
 for   J0204--1009 implies that this BCD is strongly  dominated by dark matter.

Unexpectedly, it is the most isolated BCD,  J0301--0052, that shows {one of the most asymmetric \hi\ morphologies} amongst our BCD sample. Given the isolation of J0301--0052, we would expect there to have been sufficient time for its \hi\ to have virialised and its \hi\  to be symmetrically distributed around the galaxy's optical centre, as we see in other isolated dwarfs \citep[e.g.][]{sengupta2019, Guo_24}.    {Instead, we see that the \hi\ column density maximum of J0301--0052  is projected $\sim$ 6 arcsec ($\sim$ 1 kpc) SW of the  optical centre and near the end of the optical tail. While the asymmetric \hi\ distribution could be due to a secular process, given the relative isolation of this  BCD, the \hi\ and optical tails seem more likely to be} signatures of the ongoing accretion of an \hi-rich dwarf companion.{ The argument for a tidal origin for the tail is strengthened by the DECAaLS $z$-band image, which indicates the tail includes an older stellar population; we would expect this to be disturbed during a  tidal interaction. }

{The \hi\ tail and  optical counterpart of J0315-0024  {provide clear signatures} of a recent interaction,} but it is unclear whether this is attributable to a merging satellite or an interaction with its nearest neighbour, the merging KUG\,0313-006 system. This is because of the much stronger \hi\ morphological and kinematic signatures attributable to the merger of the KUG\,0313-006 system itself, which would be expected to completely obscure any interaction signature in KUG\,0313-006 attributable to the merging {system's} interaction with J0315--0024.
 {The projected orientation of the  \hi\ tail and the optical spur tend to favour the alternative interpretation whereby the interaction signatures arise from an ongoing minor merger.}

In the case of J2053+0039, the \hi\ column density maximum is slightly offset from its optical counterpart. However, the \hi\ iso-velocity contours in Fig. \ref{fig_5_2053} show more perturbed kinematics near the centre of the optical galaxy than to the north; as noted in Sect. \ref{2053_result}, this is unlikely to be due to an interaction with SDSS J205311.31+003926.0, projected $\sim$ 23 arcsec (7 kpc)  NW of J2053+0039; see  {Figs. \ref{fig_5_2053} and \ref{fig_a1}}. We note that the velocity field at the far western side \hi\ disc of its companion, UGC\,11645,  shows a significant disturbance at the same velocities as  J2053+0039. Taken together, this is evidence of a recent tidal interaction between the BCD and UGC\,11645. This hypothesis is strengthened by the proximity of UGC\,11645 in both velocity and projected separation, as well as the fact that    J2053+0039 has the highest tidally induced \textcolor{black}{SF} parameter ($P_{gg}$ = 0.03) amongst our BCDs   (see Sect. \ref{other_samples}.)

Based on the limited resolution of even the BC-configuration VLA observations, the morphologies of three of the four BCDs present evidence ---of varying strength--- of interactions with a nearby companion within the last $\sim$ 1 Gyr.   { J0315-0024 and  J0301--0052 both present one-sided  \hi\  tails, which could plausibly be attributed to minor mergers.  Our results are therefore} consistent with the finding of perturbed \hi\ in BCDs from other \hi\ interferometric studies  \cite[e.g.][]{bravo04,Ashley17}.


\subsection{Comparison with other samples}
\label{other_samples}

\begin{figure}
\begin{center}
\includegraphics[ angle=0,scale=0.31] {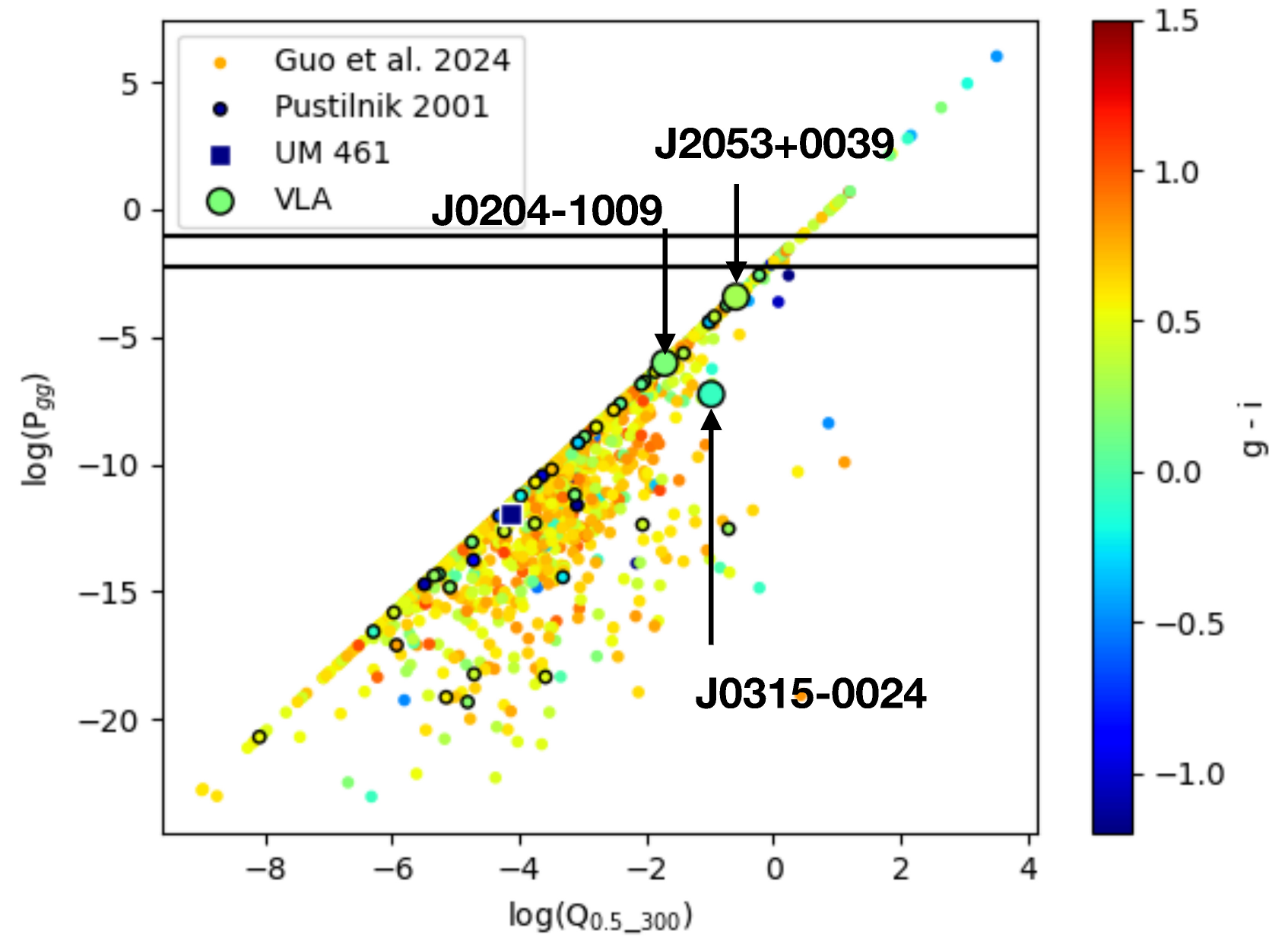}
\vspace{0.01cm}
\caption{ Comparison of $P_{gg}$, the tidally induced star formation parameter, versus Q$_{0.5-300}$, the near-neighbour tidal force parameter, between our sample (large circles with black rim) with dwarfs from \citet[][small circles no rim colour]{Guo_24},  \citet[][small circles with black rim]{Pustilnik01}, and UM\,461 \cite[][square with white rim]{lagos18}. The horizontal lines show the threshold range for tidally induced star formation  \citep{byrd90}. The colour bar shows the SDSS $g-i$ colours of the  galaxies.  }
\label{fig_12} 
\end{center}
\end{figure}

To {investigate} the effect of the local environment on the BCD properties of our sample, we carried out an analysis of the impact of tidal forces within a volume around each BCD \textcolor{black}{using the tidal force parameter $Q_{0.5-300}$  defined in \cite{Guo24a} } and compared this with a parameter used to determine whether the tidal forces attributable to the nearest neighbour are sufficient to induce 
SF in the target BCD (P$_{gg}$ \textcolor{black}{as defined in \cite{byrd90}}). These two environmental parameters were calculated using the same methods used to analyse a sample of dwarf galaxies in Sect.  4.2 of \cite{Guo24a}. The $Q_{0.5-300}$ tidal force parameter considers only companions with SDSS spectral redshifts within a radius of 500 kpc and velocities within $\pm$ 300 \km\ of that of the target BCD.  The threshold for the nearest companion to trigger \textcolor{black}{SF} depends on the characteristics of the interaction and we assume the threshold is in the $P_{gg}$ range of  0.006 to 0.1 {(log -2.2 to log -1.0) per Fig.  \ref{fig_12} \citep{byrd90}. }


Figure \ref{fig_12} shows a log scale plot of the tidal perturbation parameter P$_{gg}$  against  Q$_{0.5-300}$ for {the three BCDs in our sample with companions compared}   to samples from \citep{Pustilnik01} and the dwarf galaxy sample from \citet[n = 2568]{Guo24a}, where targets were  selected with $\log(M_*/$\msolar) $\leq$ 9. The figure also shows  UM\,461 from \citep{lagos18}. The colour scale shows the SDSS $g-i$ colour of the  galaxies. {The region between the two horizontal lines in the figure indicates the P$_{gg}$ threshold for tidally induced SF from a companion galaxy. The region above these lines is indicative of galaxies with tidally induced \textcolor{black}{SF}.} 
  In our BCD sample, one galaxy (25\%)  did not have a Q$_{0.5-300}$ companion, whereas 43\% of the members of the Guo SDSS blue metal rich dwarf sample have no companion. Pustilnik's sample is not based on SDSS data, and so instead we tried to cross-match Pustilnik's galaxies to those in the SDSS. This was successful for 57 \% of their sample. It is clear from Fig. \ref{fig_12}  that our BCDs with nearby companions (3 out of 4) come from galaxy regions with  much higher average density than the Guo's dwarfs or even Puslilnik's BCDs. {None of our BCDs are in or above the $P_{gg}$ threshold region for tidally induced \textcolor{black}{SF}, although J2053+0039 is close to the $P_{gg}$ threshold.}  \cite{Guo24a} find that dwarfs in their sample above the P$_{gg}$ threshold for tidally induced \textcolor{black}{SF} have bluer $g-i$ colours than the bulk of their sample, which they attribute to a recent tidal interaction. All three VLA BCDs with companions are bluer (mean $g-i$ = 0.12; see Table \ref{table3}) than the mean of the \cite{Guo24a} sample (0.52) and come from denser environments and are also closer to the tidally induced SF threshold than the bulk of the \cite{Guo_24} dwarf sample. The mean $g-i$ of the Pustilnik subset is also bluer than the mean $g-i$ of the \cite{Guo_24} sample,  although they span most of the same range of $\log(Q_{0.5-300})$ and $P_{gg}$  as the \cite{Guo_24} sample.  {Therefore, while none of our BCDs show a clear sign of    tidally induced \textcolor{black}{SF}, given that three of them are from higher-density regions, and given their disturbed \hi\ kinematics and elevated sSFR, we conclude that they have undergone {recent} SF-inducing interactions.} {We note that the extreme SDSS $g - i$ colour (- 1.18) of the brightest optical region of  UM\,461 is {also} consistent with a recent interaction, but the gas cloud or minor merger reported in \cite{lagos18} {could not be quantified using}  either  $\log(Q_{0.5-300})$ or $P_{gg}$  {because of the lack of a discrete} companion.}  

\begin{figure}
\begin{center}
\includegraphics[ angle=0,scale=0.26] {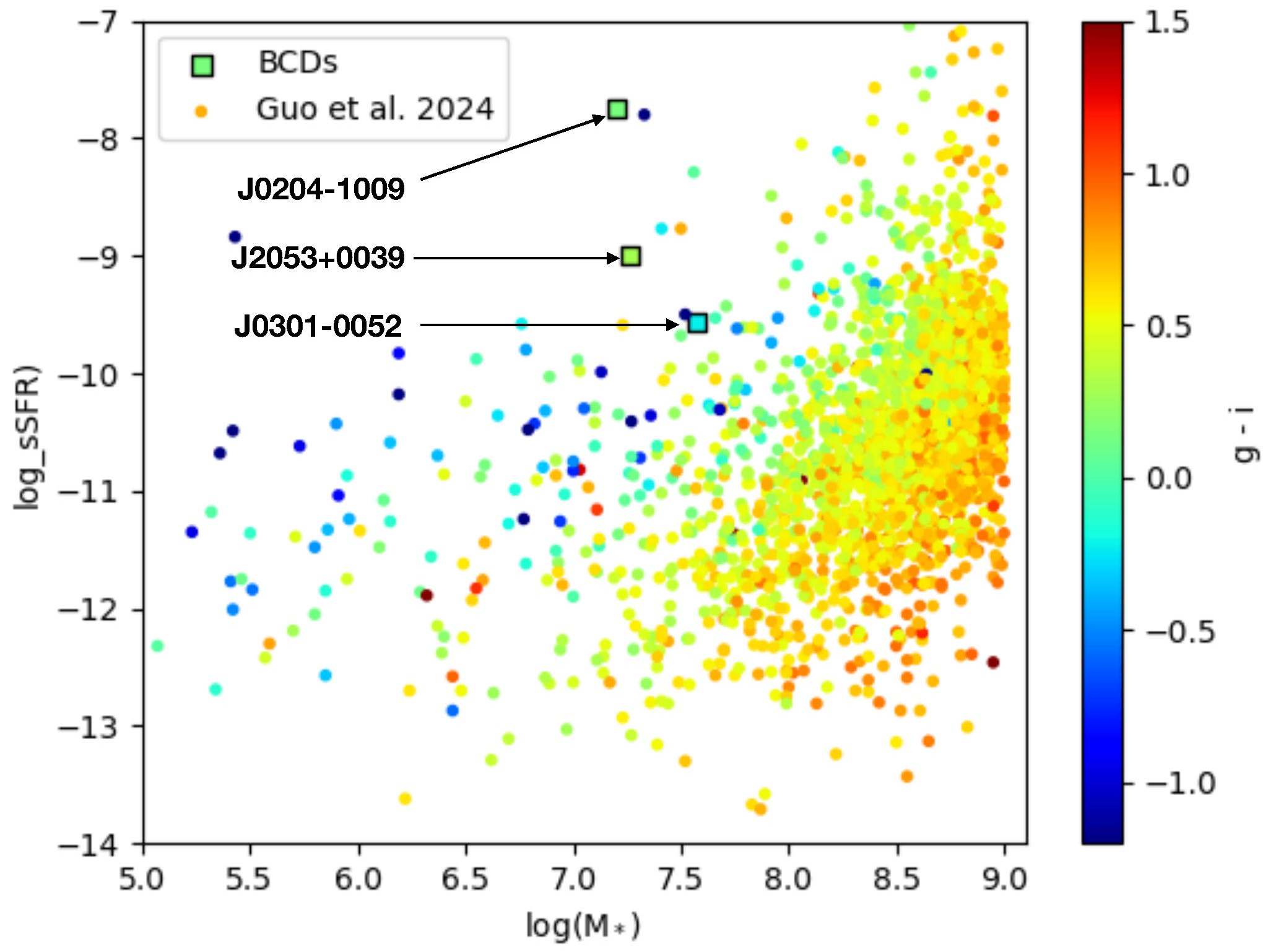}
\vspace{0.01cm}
\caption{  log(sSFR) v log(M$_*$) for three of the BCDs (square symbols) in comparison with the sample of blue metal-rich dwarf galaxies \citep[filled circles]{Guo24a}. The colour of the symbols corresponds to the SDSS $g - i$ colour of the galaxies in the colour bar on the right hand side of the figure.  }
\label{fig_14} 
\end{center}
\end{figure}

{In Fig. \ref{fig_14} we compare the log(sSFR) versus log(M$_*$) of the three BCDs of our sample that have SFRs available from \cite{filho13} with that of the sample of blue dwarfs from \cite{Guo24a}. As expected, we see from the figure that the BCDs have elevated sSFRs  compared to the Guo sample. Moreover, Figs. \ref{fig_12} and      \ref{fig_14} and Table \ref{table3} show all four BCDs have particularly blue colours, which the  \cite{Guo24a} analysis indicates is a signature of recent tidally induced \textcolor{black}{SF} in dwarf galaxies. In the case of J0301--0052, the blue SDSS $g - i$ colour (- 0.22) and sSFR in Fig. \ref{fig_14}  are consistent with an ongoing merger. The SFR and $g - i$ colours from Figs. \ref{fig_12} and \ref{fig_14}  support the conclusion from the \hi\ analysis that the elevated SFRs in our sample of BCDs are most likely attributable to recent fly-by interactions or, in the case of J0301--0052{ and possibly J0315-0024,} an ongoing merger.}

\section{Concluding remarks {and possible scenarios}}
\label{concl}

{\cite{sanchez15} argued that a low-metallicity gas-cloud impact {under} the cosmic web scenario can explain the metallicities and morphologies observed in a sample of ten cometary-like XMPs. In \cite{lagos18}, we analysed the spatial variation of 12 + log(O/H) abundances using the direct method (T$_{e}$) in the cometary-like BCD galaxy UM\,461 using optical {IFU spectroscopy}. We found that the metallicity inhomogeneities in {that}  galaxy are consistent with the recent infall of a metal-poor gas cloud in the region now exhibiting the lowest metallicity and highest \textcolor{black}{SF}.{ This infalling cloud is also visible as a perturbation in the \cite{vavzee98} \hi\ velocity field.}}

For three out of four of our BCDs {(J0204-1009, J0315-0024, and J2053+0039)}, we detect \hi\ in nearby companions.  These cases also show \hi\ morphological and/or kinematic evidence of a recent fly-by interaction with a companion, which could reasonably explain their enhanced sSFR compared to other blue star-forming dwarfs. However, it is the isolated BCD, J0301--0052, that shows the strongest cometary optical morphology. {The most straightforward explanation for {the} \hi\ and the optical cometary tails is that J0301--0052 is undergoing a merger with a gas-rich dwarf. This merger appears to be responsible for the \hi\ and optical morphologies, while at the same time triggering enhanced star formation.} {We hypothesise that, {while} cold gas accretion could produce the {\hi\ disturbances observed in void galaxies \citep{Kurapati2024a,kurapati2024}, interactions with close neighbours {or minor mergers} appear to be the main contributors to these disturbances and the enhanced sSFS} in our {samples} of XMPs \citep{lagos14,lagos16,lagos18} and those in the literature \cite[e.g.][]{filho13,sanchez15}.}

These results, {taken together with our results for UM\,461,  are consistent with the findings of \cite{Pustilnik01} based on a larger sample} that most BCDs are associated with either mild interactions or mergers. The \hi\ evidence for recent interactions of the three BCDs with nearby companions is corroborated by our analysis of the tidal forces exerted on the BCDs by companions with spectroscopic redshifts, {as well as their bluer SDSS $g - i$ colours}. In the case of the BCD J0204--1009, we have sufficient resolution to determine that the galaxy is dominated by \textcolor{black}{DM} and estimate its DM halo {mass to be} in the range of {1.2 $\times$ 10$^{11}$ to 5.2 $\times$ 10$^{11}$} \msolar, but from this sample of one we cannot form a general conclusion about the DM content of BCDs. {However, this result is consistent with the findings that other dwarf galaxies are dominated by dark matter \citep[e.g.][]{oh11}. }

{While the number of  BCDs with resolved \hi\ observations remains small, the picture  emerging from our observations is as follows: 
\begin{itemize}
\item
The presence of nearby companions, disturbed \hi\ morphology and kinematics in the BCDs and/or their companions, and enhanced blue 
$g - i$  colour in the BCDs are consistent with fly-bys or mergers triggering the enhanced SFRs seen in many BCDs.
\item
The evidence from UM\,461 and our VLA observations of J0301--0052 for mergers of either a large gas cloud or a \hi-rich dwarf explains their enhanced SFR, extremely blue $g - i$  colours, and optical cometary morphologies. J0301--0052 shares these characteristics with UM\,461. However, UM\,461 is the only one of our investigated BCDs with IFU observations and is consequently the only case where we can explain its low metallicity; we believe it to be the result of the accretion of a gas cloud with lower metallicity than the parent BCD.
\item
While our \hi\ observations show that fly-bys and mergers provide a plausible explanation for the observed sSFRs and $g - i$ colours in our sample, the case of UM\,461 demonstrates that understanding the origin of the second key characteristic of BCDs (i.e. their low metallicity) requires both resolved \hi\ and optical IFU data.
\end{itemize}}

In future work, we plan to { apply for IFU spectroscopy  observations to}  study the chemical homogeneity of the warm gas in our sample of cometary XMP BCDs. \textcolor{black}{The principal aim of these observations will be to reveal the spatially resolved origin of the low metallicity of the targets} 

\begin{acknowledgements}
PL { (DOI 10.54499/dl57/2016/CP1364/CT0010) and TS (DOI 10.54499/DL57/2016/CP1364/CT0009)} are supported by national funds through Funda\c{c}\~{a}o para a Ci\^{e}ncia e a Tecnologia (FCT) and the Centro de Astrof\'isica da Universidade do Porto (CAUP). {We wish to thank the referee for their useful comments and suggestions which have improved the paper.}    This research has made use of the NASA/IPAC Extragalactic Database (NED) which is operated by the Jet Propulsion Laboratory, California Institute of Technology, under contract with the National Aeronautics and Space Administration. This research has made use of the Sloan Digital Sky Survey (SDSS). The SDSS Web Site is http://www.sdss.org/.
This research made use of APLpy, an open-source plotting package for Python \citep{Robitaille2012}.
\end{acknowledgements}

\section*{Data availability}

The raw data for used in this article are available from the NRAO archive, and the reduced data are available at DOI 10.5281/zenodo.13997847.

\bibliographystyle{aa} 
\bibliography{cluster} 

\onecolumn
\begin{appendix}
\section{BCD DECaLS colour images }
\label{appendix_A}
\begin{figure*}[h]
\centering
\includegraphics[ angle=0,scale=0.28] {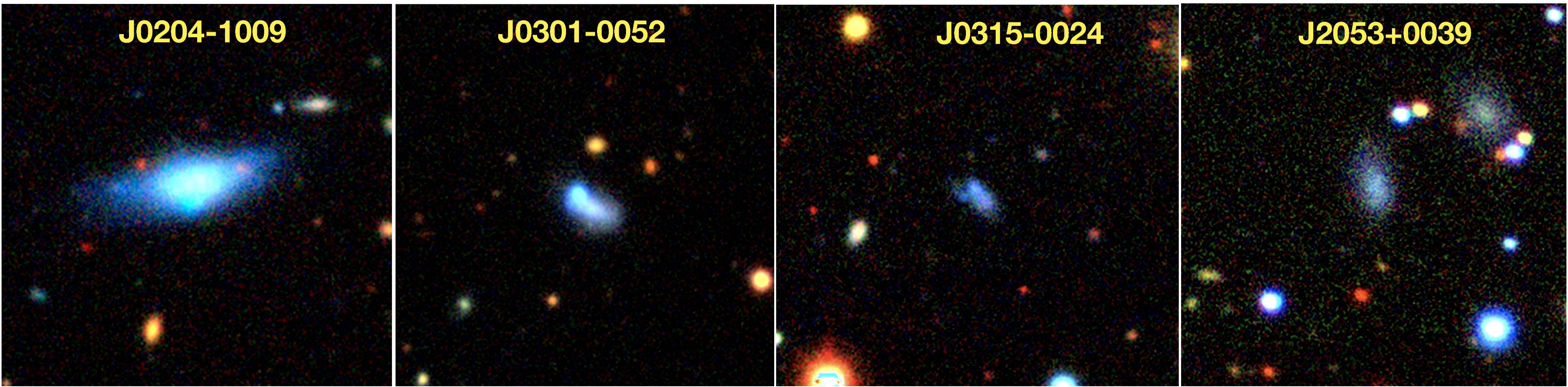}
\vspace{0.01cm}
\caption{DECaLS 1 arcmin $\times$ 1 arcmin false colour ($g$, $r$, $z$) images of our four BCD galaxies which are at the centre of each panel.}
\label{fig_a1}
\end{figure*}

{Figure \ref{fig_a1} shows 1 arcmin $\times$ 1 arcmin false colour images from DECaLS of our four BCDs. The IDs of the BCD appear at the top of each panel in the figure.} 

\section{VLA observational parameters}
\label{appendix_B}

The VLA observational parameters for each of the observed scheduling blocks are listed in Table \ref{table2}. Also given in that table are the characteristics of the data after combination of all the scheduling blocks pertaining to a single target. Synth.\ beam is the synthesised beam or point spread function taken to be the full width at half-power level (FWHP).


\begin{table}[h]
\centering
\begin{minipage}{150mm}
\caption{VLA observational parameters.}
\label{table2}
\begin{tabular}{@{}cccccrrr@{}}
\hline

Date&Target &Configuration& Int.\footnote{On target integration time.}& Cent. freq&Synth.beam&Synth bean    \\
&  &  &[hrs]&[GHz] &[arcscc]&[PA\degree]  \\
 \hline 
2020-02-02      &       J0204--1009     &       C       &       0.8     &{1.411}               \\
2020-02-04      &       J0204--1009     &       C       &       0.8     &{1.411}               \\
2020-06-28      &       J0204--1009     &       B       &       1.6     &{1.412}               \\
2020-07-08      &       J0204--1009     &       B       &       1.6     &{1.412}               \\
2020-08-08      &       J0204--1009     &       B       &       1.6     &{1.412}               \\
\bf{Combined}   &       \bf{J0204--1009}        &       \bf{BC} &       \bf{6.3}&\bf{--}&\bf{9.5 $\times$ 7.6}&\bf{-31.6}                \\
\hline

2020-02-01      &       J0301--0052     &       C       &       0.8     &{1.410}               \\
2020-02-01      &       J0301--0052     &       C       &       0.8     &{1.410}               \\
2020-07-05      &       J0301--0052     &       B       &       1.6     &{1.410}               \\
2020-07-13      &       J0301--0052     &       B       &       1.6     &{1.410}               \\
2020-07-14      &       J0301--0052     &       B       &       1.6     &{1.410}               \\
\bf{Combined}   &\bf{J0301--0052}       &       \bf{BC} &\bf{6.3}&\bf{--}&\bf{8.4 $\times$ 7.9}& \bf{-48.4}       \\
\hline
2020-01-31      &       J0315--0024     &       C       &       0.8     &{1.389}               \\
2020-02-02      &       J0315--0024     &       C       &       0.8     &{1.389}               \\
2020-07-11      &       J0315--0024     &       B       &       1.6     &{1.389}               \\
2020-07-15      &       J0315--0024     &       B       &       1.6     &{1.389}               \\
2020-07-20      &       J0315--0024     &       B       &       1.6     &{1.389}               \\
\bf{Combined}   &       \bf{J0315--0024}        &       \bf{BC}&\bf{6.3}&\bf{--}&\bf{9.5 $\times$ 8.3}&\bf{-32.2}        \\
\hline
2020-02-02      &       J2053+0039      &       C       &       0.8     &{1.402}               \\
2020-02-02      &       J2053+0039      &       C       &       0.8     &{1.402}               \\
2020-06-29      &       J2053+0039      &       B       &       1.6     &{1.402}               \\
2020-06-30      &       J2053+0039      &       B       &       1.6     &{1.402}               \\
2020-07-03      &       J2053+0039      &       B       &       1.6     &{1.402}               \\
\bf{Combined}   &       \bf{J2053+0039}\footnote{{For the J2053+0039 field, cubes were made at two different spacial resolutions for the BCD  (7.9" $\times$ 6.6")  and UGC 11645 (12.4" $\times$ 9.3"), see Sect. \ref{2053_result}.} }   &       \bf{BC} &       \bf{6.3}&\bf{--}&\bf{12.4 $\times$ 9.3}   &\bf{31.7}
        \\
\bf{Combined}   &       \bf{J2053+0039} &       \bf{BC} &       
\bf{6.3}&\bf{--}&\bf{7.9 $\times$ 6.6}  &\bf{31.7}
        \\
 \hline
\end{tabular}
\end{minipage}
\end{table}

\newpage
\section{VLA BC configuration \hi\ spectra}
\label{appendix_C}
\begin{figure*}[h]
\centering
\includegraphics[ angle=0,scale=0.37] {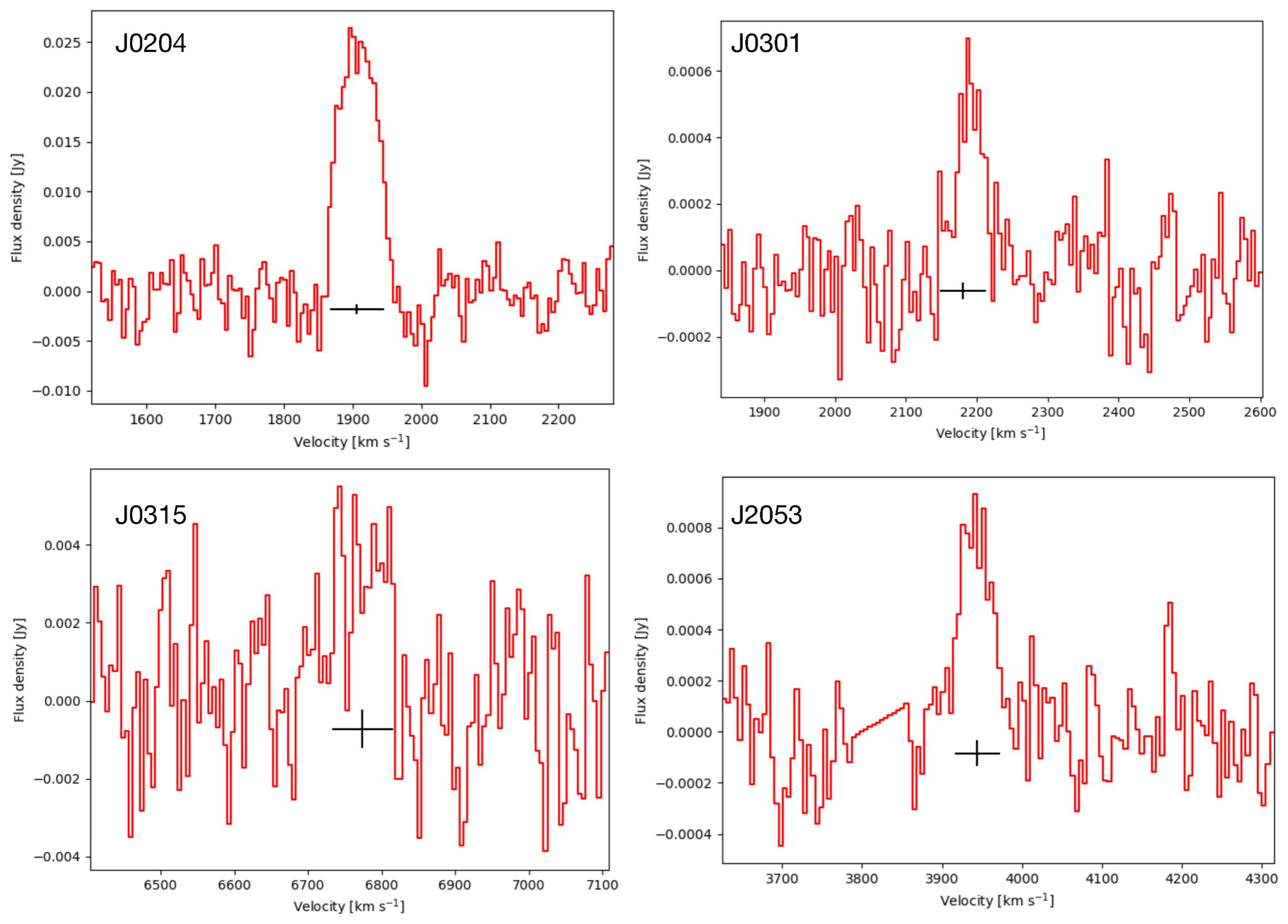}
\vspace{0.01cm}
\caption{Spectra of the VLA BC-configuration BCD detections at a velocity resolution of 5\,\km. The horizontal line over-plotted below each spectrum indicates the W$_{20}$ velocity range. The black vertical lines on the W$_{20}$  line are the V$_{HI}$ calculated,   using the V$_{sys_w20}$ method from \citep{Reynolds21}.  V$_{HI}$ (V$_{sys_w20}$ method) and   W$_{20}$ measurements for each galaxy in this figure are shown in Table \ref{table3}. }
\label{fig_6}
\end{figure*}

The spectra of the BCDs are shown in Fig. \ref{fig_6}.

\end{appendix}
\label{lastpage}
\end{document}